\documentclass[10pt,pra,aps,twocolumn,superscriptaddress,showpacs]{revtex4-1}
\usepackage{amsmath,amsfonts,amssymb}
\usepackage{braket} 
\usepackage{ulem} 
\usepackage{framed} 
\usepackage[pdftex]{graphicx}
\usepackage{natbib}
\usepackage{bbm}    
\usepackage{color}
\usepackage{multirow} 

\newcommand{\bs}[1]{\boldsymbol{#1}}

\begin{document}

\title{Coulomb drag between helical Luttinger liquids}

\author{N. Kainaris}
\affiliation{Institut f\"ur Nanotechnologie, Karlsruhe Institute of Technology, 76021 Karlsruhe, Germany}
\affiliation{\mbox{Institut f\"ur Theorie der Kondensierten Materie, Karlsruher Institut f\"ur
  Technologie, 76128 Karlsruhe, Germany}}

\author{I. V. Gornyi}
\affiliation{Institut f\"ur Nanotechnologie, Karlsruhe Institute of Technology, 76021 Karlsruhe, Germany}
\affiliation{\mbox{Institut f\"ur Theorie der Kondensierten Materie, Karlsruher Institut f\"ur Technologie, 76128 Karlsruhe, Germany}}
\affiliation{A.F. Ioffe Physico-Technical Institute, 194021 St.~Petersburg, Russia}
\affiliation{L.D. Landau Institute for Theoretical Physics, 119334 Moscow, Russia}

\author{A. Levchenko}
\affiliation{Department of Physics, University of Wisconsin-Madison, Madison, Wisconsin 53706, USA}

\author{D. G. Polyakov}
\affiliation{Institut f\"ur Nanotechnologie, Karlsruhe Institute of Technology, 76021 Karlsruhe, Germany}


\begin{abstract}
We theoretically study Coulomb drag between two helical edges with broken spin-rotational symmetry,
such as would occur in two capacitively coupled quantum spin Hall insulators. For the helical edges,
Coulomb drag is particularly interesting because it specifically probes the inelastic interactions
that break the conductance quantization for a single edge. Using the kinetic equation formalism,
supplemented by bosonization, we find that the drag resistivity $\rho_D$ exhibits a nonmonotonic
dependence on the temperature $T$. In the limit of low $T$, $\rho_D$ vanishes with decreasing $T$
as a power law if intraedge interactions are not too strong. This is in stark contrast to Coulomb
drag in conventional quantum wires, where $\rho_D$ diverges at $T\to 0$ irrespective of the strength
of repulsive interactions. Another unusual property of Coulomb drag between the helical edges concerns
higher $T$ for which, unlike in the Luttinger liquid model, drag is mediated by plasmons.
The special type of plasmon-mediated drag can be viewed as a distinguishing feature of the helical
liquid---because it requires peculiar Umklapp scattering only available in the presence of a Dirac
point in the electron spectrum.
\end{abstract}

\maketitle

\section{Introduction}
\label{sec:introduction}

The helical Luttinger liquid (HLL) emerges at the edge of a two-dimensional quantum spin Hall (QSH)
insulator~\cite{kane_mele_2005,Kane_Mele_2005(2),bernevig_2006,hasan10,qi11,maciejko11} and consists,
in its most conventional form protected by time-reversal symmetry, of two counterpropagating Kramers
conjugate modes. In an ``ideal" helical edge, the electron spin is conserved for each of the chiral modes
(``$S_z$-conserving models"). Electron-electron backscattering between the modes, as well as backscattering
by nonmagnetic inhomogeneities, is then prohibited by the combination of the spin-axial and time-reversal
symmetries. As a consequence, charge transport through the ideal helical edge is characterized by a
quantized conductance $G_0 = 2 e^2/h$, independent of the temperature $T$, also in the presence of
nonmagnetic disorder. Experimentally, the conductance quantization has been observed at the edges of
HgTe/CdTe~\cite{Koenig_2007,Roth_2009} and InAs/GaSb~\cite{Knez_2011,Knez_2014,Du_2015} quantum wells.

In a more realistic description of topological insulator materials, spin-rotational invariance is not
preserved in the helical edge.
One perturbation that violates the invariance is Rashba-type spin-orbit coupling induced by broken
inversion symmetry about the plane of the semiconductor heterostructure. In the presence of nonmagnetic
disorder, {\it elastic} backscattering between Kramers partners remains exactly forbidden by time-reversal
symmetry, irrespective of the presence or absence of spin-rotational invariance. As a result,
the $T=0$ conductance is given by $G_0$ independently of the strength of disorder (as long as the
two-dimensional bulk is insulating). However, at nonzero $T$, {\it inelastic} backscattering is generically
triggered beyond the $S_z$-conserving models~\cite{Schmidt_2012} and gives rise to dissipation, even in the
absence of disorder, modifying significantly the transport properties of both a clean and disordered helical
liquid \footnote{Reference \cite{Schmidt_2012}, while studying inelastic backscattering in a helical
edge with broken spin-axial symmetry, admits
the existence of inelastic backscattering also in models that preserve this symmetry. We believe, however,
that inelastic backscattering in a helical edge is only possible if the spin-locking axis changes its orientation
with varying $k$. In particular, the earlier works~\cite{Kane_Mele_2005(2),Wu_2006} cited in
Ref.~\cite{Schmidt_2012} also tacitly rely on broken spin-axial symmetry in the context of a nonzero
backscattering rate.}. In the limit of low $T$, inelastic backscattering leads to $T$ dependent corrections
to the quantized edge conductance~\cite{Schmidt_2012,Lezmy_2012,Budich_2012,Crepin_2012,Geissler_2014}. These behave, generically, as power
laws of $T$, similar to the conventional Tomonaga-Luttinger liquid (TLL). In the thermodynamic limit,
the corrections to the conductance convert into a finite conductivity
\cite{Strom_2010,Kainaris_2014,Chou_2015,Xie_2016} that is a power-law function of $T$~\footnote{The emergence of a
nonzero resistivity in a helical liquid at finite $T$, because of \mbox{Umklapp} scattering,
was mentioned in Ref.~\cite{Kane_Mele_2005(2)}}. Strong repulsive interactions are expected to block
zero-$T$ transport through the edge by spontaneously breaking time-reversal symmetry and gapping the edge
modes~\cite{Wu_2006}. Experimental evidence pointing towards the importance of the TLL effects in transport
through a strongly interacting HLL has recently been reported~\cite{Du_2015(2)}.

In this paper, we investigate Coulomb drag between parallel clean (no disorder) edges of two identical
QSH insulators, brought in proximity to each other, in the case of broken spin-rotational invariance.
We assume that each of the edges consists of a single pair of Kramers partners with a linear dispersion
relation. In a Coulomb drag measurement, current is driven in an ``active'' conductor (active edge in our setup),
inducing an electrical field or current in a ``passive'' conductor (passive edge), with the frictional force being
due to electron-electron interactions, without transfer of electrons between the subsystems.
As such, Coulomb drag is a sensitive probe of inelastic electron-electron scattering. For helical edges,
Coulomb drag is particularly worthy of study because, as already mentioned above, inelastic electron-electron
scattering is the only source of dissipation inside a single edge.

The key quantity characterizing friction is the drag resistivity,
\begin{align}
\rho_D = - E_2/j_1~,
\label{1}
\end{align}
where $j_1$ is the current density driven in the active conductor and $E_2$ is the electric field applied to
the passive conductor to compensate for the friction force and maintain zero current therein. In conventional
one-dimensional (1D) systems (single-channel quantum wires), Coulomb drag has been intensively studied both
theoretically~\cite{Flensberg_1998,Nazarov_Averin_1998,Ponomarenko_Averin_2000,Klesse_Stern_2000,Komnik_Egger_2001,Trauzettel_2002,Pustilnik_2003,
Schlottmann_2004, Fuchs_2005, Fiete_2006, Peguiron_2007, Aristov_2007,rozhkov08,Dmitriev_Gornyi_Polyakov_2012,dmitriev16}
and experimentally~\cite{Debray_2000,Debray_2001,Debray_2002,Yamamoto_2002,Yamamoto_2006,Laroche_2011,Laroche_2014}.
In general, Coulomb drag in one dimension, independently of the particular form of the electron dispersion relation,
can only occur in the presence of scattering that changes the chirality of electrons~\cite{Dmitriev_Gornyi_Polyakov_2012}.
Therefore, there is no Coulomb drag between clean ideal HLLs, in which scattering between left- and right movers is
strictly prohibited. Conversely, Coulomb drag between HLLs is only possible if there is a perturbation that breaks
spin-rotational invariance of the edge modes.

One possibility to break the axial spin symmetry and lift the restriction on backscattering interactions between
two helical liquids is to apply an external magnetic (Zeeman) field (the magnetic field also breaks time reversal
symmetry and generically gaps the edge modes). Coulomb drag between helical liquids in the presence of a magnetic
field $h$ perpendicular to the spin-locking axis was studied in Ref.~\cite{Zyuzin_Fiete_2010}.
Above the temperature at which a zigzag-ordered charge density wave is formed (and for sufficiently small $h$,
so that the electron spectrum can be approximated as linear), Ref.~\cite{Zyuzin_Fiete_2010} found
$\rho_D \propto h^4 T^{4K_- - 3}$, where $K_-$ is the Luttinger constant for the relative charge mode.
The power-law exponent of the $T$ dependence is here the same as for Coulomb drag between spinless TLLs,
but $\rho_D$ is strongly suppressed, compared to the TLL case, in the limit of small $h$.

Another possibility to destroy the spin-rotational invariance originates from spin-orbit coupling in the bulk of
the spin Hall insulator. This is the model that we study in this paper, within the framework introduced in
Ref.~\cite{Schmidt_2012} for a single helical edge. Of central importance to us is that, in contrast to
the magnetic field-induced drag~\cite{Zyuzin_Fiete_2010}, time-reversal symmetry is preserved in this model,
so that the topological nature of the edge states as Kramers partners remains intact. The significance of this
distinction is that Coulomb drag between helical liquids, if it is facilitated by spin-orbit coupling, differs in
an essential way both from Coulomb drag induced by the Zeeman field and from Coulomb drag between TLLs.

Regarding this distinction, two most important points to emphasize are the following. First, the strong suppression
of electron-electron backscattering in the limit of low $T$ makes the formation of a strong-coupling ground state,
which is a hallmark of the TLL with repulsive interactions (where a zigzag-ordered charge density wave is inevitably
formed in the low-$T$ limit), only possible if the strength of interactions exceeds a critical value.
Below the critical value, $\rho_D$ vanishes with decreasing $T$, in contrast to the TLL case. Second, time-reversal
symmetry necessitates the existence of a Dirac point in the HLL spectrum. In the vicinity of the Dirac
point, \mbox{Umklapp} scattering contributes to Coulomb drag in addition to backscattering. This results in the
emergence of a peculiar regime of plasmon-mediated Coulomb drag at higher $T$.

To make a systematic treatment of the peculiarities of Coulomb drag between helical liquids, we formalize our approach
from two complementary perspectives. We first study the kinetic equation for weakly interacting helical fermions.
Having established---for the case of weak interactions---the vanishing of $\rho_D$ in the limit of low $T$ and the
dominance of plasmon-mediated drag for higher $T$, we proceed to bosonize the model and include the TLL renormalization
effects using the Kubo formula.

The paper is organized as follows. In Sec.~\ref{Sec:Model for helical liquids with broken spin-rotational invariance},
we formulate the model of two capacitively coupled helical edges with broken spin-rotational symmetry.
Section~\ref{Sec:Coulomb drag in helical edges} is devoted to the study of Coulomb drag between weakly-interacting
edges within the kinetic equation approach. We write down the kinetic equation for two-particle scattering in the helical
edges in Sec.~\ref{Sec:Kinetic equation for capacitively coupled helical edges}. The high- and low-frequency regimes of
Coulomb drag are considered in Secs.~\ref{Sec:High frequency drag} and \ref{Sec: Low frequency drag}, respectively.
In Sec.~\ref{sIIIc}, we discuss dynamically screened interaction in the helical edges. In Secs.~\ref{sIIId} and \ref{sIIIe},
we obtain, respectively, the electron-hole and plasmon contributions to the drag rate. In Sec.~\ref{LL-boson},
we address the effects of strong intraedge interaction on the drag resistivity within the bosonization framework.
The renormalization of the first- and second-order backscattering amplitudes is analyzed in Secs.~\ref{s4}
and \ref{Sec:Renormalization group analysis}, respectively. Section~\ref{Sec:Luttinger liquid renormalization} deals
with the renormalization of the drag resistivity. The strong coupling regime is discussed in Sec.~\ref{Sec:Strong coupling drag}.
Section~\ref{Sec: Summary} concludes with a summary. Some of the technical details are moved to the Appendices.

Throughout, we use the abbreviation $\int_k = \int \! \textstyle{ \frac{\mathrm{d} k}{2 \pi} }$.

\section{The model}
\label{Sec:Model for helical liquids with broken spin-rotational invariance}

\begin{figure}
  \centering
  \includegraphics[width=.3\textwidth]{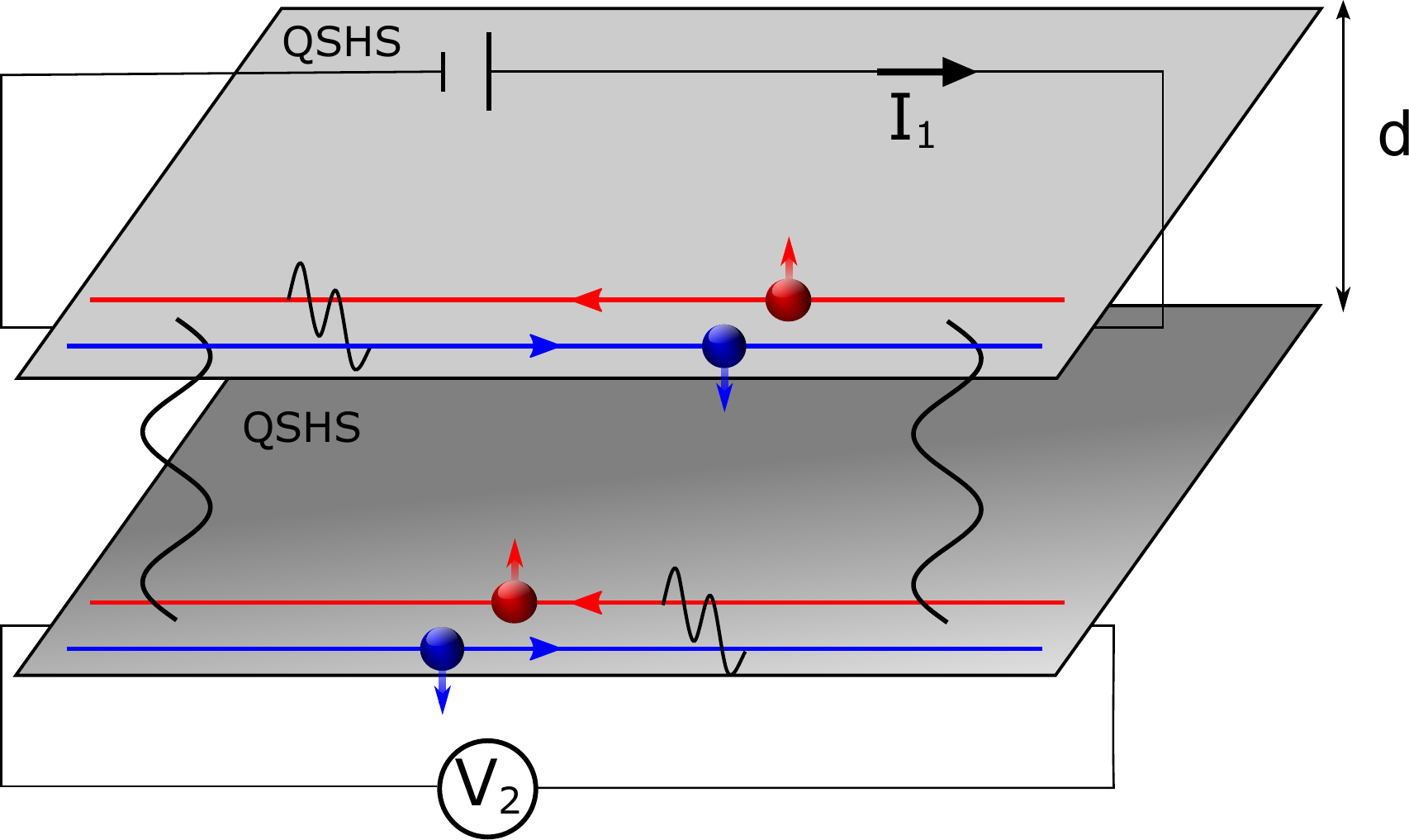}
  \caption{Schematics of a Coulomb drag measurement between helical edges of two QSH systems.
  Current $I_1$ is driven through the active edge and, as a result of electron-electron interactions,
  voltage $V_2$ is induced in the passive edge.}
  \label{Fig:dragsetup}
\end{figure}

We start by formulating our model for two helical liquids with broken spin-rotational invariance
coupled by a screened Coulomb interaction. In substance, we employ the model proposed---for a single
helical edge---in Ref.~\cite{Schmidt_2012} and extend it to the case of two edges.
We consider two identical QSH systems at the same chemical potential $\mu$, each with one Kramers pair
at the edge, as shown in Fig.~\ref{Fig:dragsetup}. Tunneling between the two QSH systems is neglected.
The transverse size of the edge channels is assumed to be much smaller than the interedge distance $d$.
We focus on Coulomb drag between infinitely long edges, i.e., the wavevector of the external perturbation
in the response functions is sent to zero before taking the dc limit: this is the order of limits that
defines the dc resistivity in general, and the dc drag resistivity $\rho_D$ in particular.

The kinetic part $H_0$ of the Hamiltonian is given by
\begin{align}
H_0=\sum_{\sigma\eta}\int_k\,(v\eta k -\mu)\,\psi_{k\sigma\eta}^\dagger\psi_{k\sigma\eta}~,
\label{kineticpart}
\end{align}
where $\psi_{k\sigma\eta}$ is the electron operator at the momentum $k$ in edge $\sigma=1,2$
with the chirality $\eta=\pm$, and $v$ is the velocity in the linear dispersion relation.
The sum over $k$ for each of the chiralities runs from $-\infty$ to $\infty$ (the bandwidth of
the edge states is assumed to be larger than all other relevant energy scales), with the chiral
spectral branches crossing at $k=0$ (``Dirac point'').

In the ideal helical edge, the spin-locking axis is independent of $k$, so that the electron state
$\tilde\psi_{k\sigma s}$ with a given spin projection $s=\uparrow,\downarrow$ coincides with $\psi_{k\sigma\eta}$.
As already discussed in Sec.~\ref{sec:introduction}, we specialize to the model in which the spin-rotational invariance
of the helical edges is broken by Rashba-type spin-orbit coupling in the bulk. In the absence of spin-axial
symmetry, $\tilde\psi_{k\sigma s}$ is generically a mixture of the chiral states $\psi_{k\sigma\eta}$ with both
chiralities $\eta$. By time-reversal symmetry, the unitary transformation between the two basis
sets (``spin'' vs ``chiral'') in the vicinity of the Dirac point has a universal form, as far as the dependence
on $k$ is concerned, to order ${\cal O}(k^2)$. Specifically~\cite{Schmidt_2012},
\begin{align}
\begin{split}
\tilde\psi_{k\sigma\uparrow}\simeq\psi_{k\sigma +}-\frac{k^2}{k_0^2}\psi_{k\sigma -}~,\\
\tilde\psi_{k\sigma\downarrow}\simeq\psi_{k\sigma -}+\frac{k^2}{k_0^2}\psi_{k\sigma +}~,
\label{helicalrotation}
\end{split}
\end{align}
where $k_0$ is a model-dependent momentum scale which characterizes the strength of spin-orbit
coupling (taken to be identical in the two edges). We assume that the spin-orbit interaction is a
weak symmetry-breaking perturbation with $vk_0\gg\max\{|\mu|,T\}$, so that the quadratic-in-$k$
expansion (\ref{helicalrotation}) is sufficient for our purposes (here and below, we count $\mu$
from the Dirac point).

The density-density interaction term in the Hamiltonian is then written in the chiral basis,
rotated with respect to the spin basis according to Eq.~(\ref{helicalrotation}), as
\begin{align}
\begin{split}
H_{\text{int}}&=\frac{1}{2}\sum_{\sigma\sigma'\eta_1\eta_2\eta_3\eta_4}\int_{kk'q}\!b_{\eta_1\eta_4}(k+q,k)b_{\eta_2\eta_3}(k'-q,k') \\
& \times V_{\sigma\sigma'}(q)\,\psi_{k+q,\sigma\eta_1}^{\dagger}\psi_{k'-q,\sigma'\eta_2}^{\dagger}\psi_{k'\sigma'\eta_3}\psi_{k\sigma\eta_4}~,
\label{model}
\end{split}
\end{align}
where $V_{\sigma\sigma'}(q)$ is the Fourier component of the interaction potential inside
($\sigma=\sigma'$ equal to 1 or 2) and between ($\sigma\neq\sigma'$) the edges and
\begin{align}
b_{\eta_1\eta_2}(k_1,k_2)=\delta_{\eta_1\eta_2}-\eta_1\delta_{\eta_1,-\eta_2}\frac{k_1^2 -k_2^2}{k_0^2}~.
\label{elementsB}
\end{align}
We assume that the interactions in the double-edge system are screened by a nearby metallic gate.
Note that---irrespective of the relation between the distance to the gate and the distance between
the wires $d$---the interwire potential $V_{12}(q)$ starts to decay exponentially with
increasing $|q|$ at $|q|\sim 1/d$ (see, e.g., Appendix A of Ref.~\cite{Dmitriev_Gornyi_Polyakov_2012}).
For simplicity, we take $V_{11}(q)=V_{22}(q)$ to be given by a constant $V_{0\parallel}$ and  $V_{12}(q)$
by a simple exponential $V_{0\perp} e^{-|q| d}$. Throughout Sec.~\ref{Sec:Coulomb drag in helical edges},
we also assume that $V_{0\parallel}=V_{0\perp}$.

The presence of the factors (\ref{elementsB}) in the interacting part of the Hamiltonian for the helical
liquid constitutes the key difference between the helical and conventional Luttinger models.
Another difference to notice is related to the population of the eigenstates at thermal equilibrium.
The conventional Luttinger model is formulated for $T\ll |\mu|$, where the chemical potential $\mu$ is
counted from the energy at which the chiral spectral branches meet [either at the bottom of the electron
spectrum, linearized in the vicinity of the Fermi energy, or at the crossing point of two chiral branches
with a linear dispersion relation, similarly to Eq.~(\ref{kineticpart})]. In the helical Luttinger model,
we assume that $T$ can be larger than the energy difference between the Fermi level and the Dirac point,
so that the real scattering processes that involve the electron states at and around the Dirac point are
not necessarily thermally suppressed.

\section{Coulomb drag between helical edges: Kinetic theory}
\label{Sec:Coulomb drag in helical edges}

As mentioned in Sec.~\ref{sec:introduction}, we first consider Coulomb drag between helical edges within
the kinetic equation framework in the limit of weak interactions, by neglecting the TLL renormalization of
the parameters of the system. This is the same limit that was studied in Ref.~\cite{Schmidt_2012}
for a single edge. As will be seen below, the double-edge system for the case of weak interactions
exhibits an essentially richer behavior---as far as the transport mechanisms are concerned---than the
single edge, primarily because of an important subtlety in the plasmon-mediated coupling between the edges.

\subsection{Kinetic equation}
\label{Sec:Kinetic equation for capacitively coupled helical edges}

The kinetic equation for $f_\sigma$, the distribution function in edge $\sigma$, reads
\begin{align}
\partial_t f_{\sigma}(1) - e E_{\sigma} \partial_{k_1}f_{\sigma}(1) = \text{St}_{\sigma}(1)~,\label{KE}
\end{align}
where $E_\sigma$ is the electric field ($e>0$) in edge $\sigma$ and the argument of $f_\sigma(1)$
includes both the momentum and chirality, with 1 being a shorthand notation for $k_1$ and $\eta_1$, etc.
As a starting point, we neglect intraedge collisions (these will be included in
Sec.~\ref{Sec: Low frequency drag}) and write the collision integral ${\rm St}_\sigma (1)$ for pair
collisions as (for $\sigma=1$)
\begin{widetext}
\begin{align}
\begin{split}
\text{St}_{1}(1) =& \;  (2 \pi)^2 \sum_{\eta_2\eta_{1'}\eta_{2'}}\int_{k_2k_{1'}k_{2'}} \!
\big|V_{12}(k_{1'}-k_1,\epsilon_{1'}-\epsilon_1)\big|^2 |b_{\eta_{1'}\eta_1}(k_{1'},k_1)|^2 |b_{\eta_{2'}\eta_2}(k_{2'},k_2)|^2\,
\delta(k_1+k_2-k_{1'}-k_{2'}) \\ & \times \delta(\epsilon_1+\epsilon_2-\epsilon_{1'}-\epsilon_{2'})\,
\Big\lbrace f_1(1') f_2(2') [1-f_1(1)] [1-f_2(2)]
-f_1(1) f_2(2) [1-f_1(1')] [1-f_2(2')]     \Big\rbrace \label{CI} \,
\end{split}
\end{align}
where $\epsilon_1 = v\eta_1k_1$, etc. The collision integral ${\rm St}_2(1)$ for electrons in edge 2
is obtained by exchanging the edge indices $1 \leftrightarrow 2$ of the distribution functions.
The dynamically screened RPA interaction $V_{12}(q,\Omega)$, whose derivation is given in Appendix~\ref{a1},
is specified in Eq.~(\ref{RPAV12}) below. As will be shown in Sec.~\ref{sIIIe}, screening plays a crucial
role in the present problem for not too low $T$ as it opens up a peculiar plasmon-mediated scattering channel
for Coulomb drag.

It is convenient to represent $f_\sigma (1)$ in terms of the function $g_{\sigma}(1)$ as
\begin{align}
f_{\sigma}(1) = n_F(\epsilon_1) -g_{\sigma}(1) n_F(\epsilon_1) [1-n_F(\epsilon_1)] \, ,
\end{align}
where $n_F(\epsilon_1) = [1+\exp(\epsilon_1 - \mu)/T]^{-1}$ is the thermal distribution function.
Linearizing Eq.~(\ref{KE}) in $g_{\sigma}$, we obtain (in the $\omega$ representation)
\begin{align}
-i\omega g_{\sigma}(1)-\eta_1\frac{eE_{\sigma}v}{T}=\text{st}_{\sigma}(1) \label{KE1}~,
\end{align}
where (for $\sigma=1$)
\begin{multline}
\text{st}_{1}(1)
= {4\over \zeta^2(1)}\,(2 \pi)^2  \sum_{\eta_2\eta_{1'}\eta_{2'}}\int_{k_{2}k_{1'}k_{2'}} \,
\big|V_{12}(k_{1'}-k_1,\epsilon_{1'}-\epsilon_1)\big|^2
\big|b_{\eta_{2'}\eta_2}(k_{2'},k_2)
\big|^2 \big|b_{\eta_{1'}\eta_1}(k_{1'},k_1)\big|^2
\vphantom{\big|b_{\eta_{2'},\eta_2}(k_{2'},k_2)
\big|^2} \delta(\epsilon_1+\epsilon_2-\epsilon_{1'}-\epsilon_{2'})\\
\times \delta(k_1+ k_2-k_{1'}-k_{2'})\, \vphantom{\big|B_{\eta_{2'},\eta_2}(k_{2'},k_2)
\big|^2}  n_F(\epsilon_1)n_F(\epsilon_2)[1-n_F(\epsilon_{1'})][1-n_F(\epsilon_{2'})]
\vphantom{\big|b_{\eta_{2'},\eta_2}(k_{2'},k_2)
\big|^2} \left[\, g_1(1') + g_2(2') - g_1(1) -g_2(2) \,\right]\, .
\label{CI2}
\end{multline}
and
\begin{align}
\zeta(1) = \frac{1}{\cosh[(\epsilon_1-\mu)/2T]}~.
\end{align}
The electric current in edge $\sigma$, as a linear response to the fields $E_\sigma$, is related to
the solution of Eq.~(\ref{KE1}) by
\begin{align}
\begin{split}
j_{\sigma}=\frac{ev}{4}\sum_{\eta_1}\eta_1\!\int_{k_1}\!\zeta^2(1)\,g_{\sigma}(1)~.
\label{current}
\end{split}
\end{align}
The dc drag resistivity $\rho_D$ is conventionally defined in terms of $j_\sigma$ and $E_\sigma$ as
in Eq.~(\ref{1}). For the discussion in Sec.~\ref{Sec:High frequency drag}, we introduce also the
$\omega$-dependent drag conductivity defined as
\begin{align}
\sigma_{12}=j_1/E_2
\label{ac_drag_cond}
\end{align}
under the condition that $E_1=0$.

\subsection{High-frequency Coulomb drag: Scattering rate}
\label{Sec:High frequency drag}

In the limit of large $\omega$, Eq.~(\ref{KE1}) can be solved for $g_\sigma(k,\eta)$ iteratively by
expanding the solution in powers of $1/\omega$. Neglecting collisions between particles (${\rm st}_\sigma\to 0$) gives
\begin{align}
g_{\sigma}^{(0)}(k,\eta) = \frac{1}{-i \omega+0}\,\eta\,\frac{e E_{\sigma}v}{T}~.
\label{g0}
\end{align}
By substituting Eq.~(\ref{g0}) in ${\rm st}_\sigma$, the dissipative part of $\sigma_{12}$
[Eq.~(\ref{ac_drag_cond})] for large $\omega$ is then obtained, to order $1/\omega^2$, as
\begin{equation}
\text{Re}\,\sigma_{12}\simeq -\frac{e^2 v}{\pi \omega^2 \tau_D^{\infty}}, \qquad
\omega \tau_D^{\infty} \gg 1~,
\label{sigma-inf}
\end{equation}
where
\begin{align}
\begin{split}
\frac{1}{\tau^{\infty}_D} = &-\frac{(2 \pi)^3}{T}\sum_{\eta_1\eta_2\eta_{1'}}  \int_{k_1k_2k_{1'}k_{2'}}\,
\eta_1 \eta_2 \; \big|V_{12}(k_{1'}-k_1,\epsilon_{1'}-\epsilon_1)\big|^2
\big|b_{-\eta_{2},\eta_2}(k_{2'},k_2)\big|^2
\big|b_{\eta_{1'}\eta_1}(k_{1'},k_1)\big|^2  \\
& \times \vphantom{\frac{1}{1}} \delta [\,\eta_1 k_1+ \eta_2 (k_2+k_{2'})-\eta_{1'} k_{1'}] \,
\delta(k_1+ k_2-k_{1'}-k_{2'})   \,n_F(\epsilon_1)n_F(\epsilon_2)[1-n_F(\epsilon_{1'})][1-n_F(-v\eta_2k_{2'})] \, .
\end{split}
\end{align}
The symbol $\infty$ here is used to emphasize that the ``drag rate" $1/\tau_D^\infty$ is calculated in
the high-frequency limit. Importantly, since $g_{\sigma}^{(0)}$ is independent of $k$, backscattering of
at least one particle involved in the collision process is required to produce a nonzero drag rate.
Specifically, the $1/\tau_D^\infty$ is a sum of contributions of four scattering channels:
\begin{align}
\begin{array}{rlllrlll}
\text{(ai)} & \eta_1 = \eta_{1'} & \text{and} & \eta_2 = \eta_{1}~, \qquad &  \text{(bi)} &
\eta_1 =- \eta_{1'} & \text{and} & \eta_2 = -\eta_{1}~, \\
\text{(aii)} & \eta_1 = \eta_{1'} & \text{and} & \eta_2 =-\eta_{1}~, \qquad &  \text{(bii)} &
\eta_1 =- \eta_{1'} & \text{and} & \eta_2 = \eta_{1}~. \label{scatteringchannels}
\end{array}
\end{align}
The different scattering channels are depicted in Fig.~\ref{Fig:scatteringprocesses}
(together with their g-ology classification).

\begin{figure*}
  \centering
  \includegraphics[width= \textwidth]{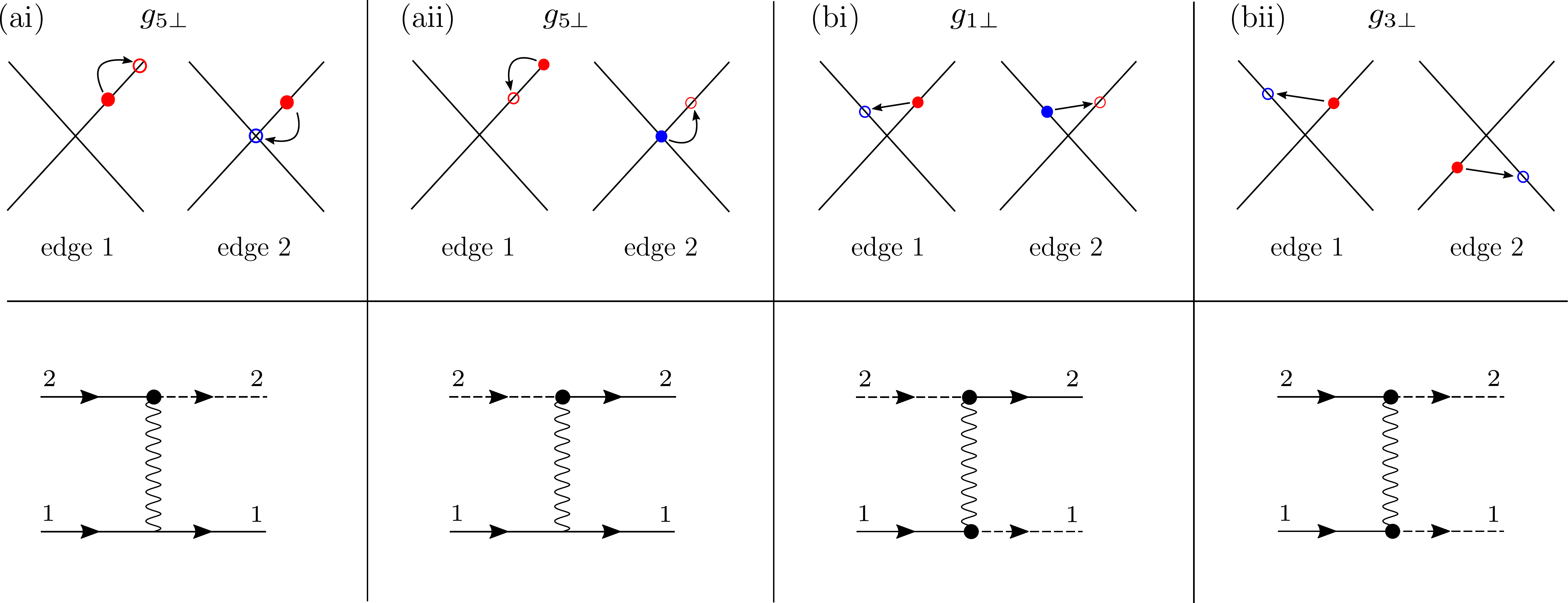}
  \caption{Momentum configurations for two-particle scattering in channels (a) and (b)
  [defined in Eq.~(\ref{scatteringchannels})] and the corresponding g-ology classification.
  The first row depicts the electron spectrum in edges 1 and 2.
  Initial and final states are shown as filled and empty circles, respectively.
  Blue (red) circles denote left (right) movers.
  The second row shows the diagrams that correspond to the scattering mechanisms above them. Here, the solid (dotted) lines refer to the quasiparticle Green's functions of right (left) movers in edge $1$ or $2$ and the thick dots denote the offdiagonal elements of the vertex function $b_{\eta_1\eta_2}(k_{\text{in}},k_{\text{out}})$ defined in Eq.~(\ref{elementsB}). 
    For the $g_{5\perp}$ processes (channel a),
  the left mover is at zero energy while the right movers are close to the Fermi surface.
  As explained in the text, the contributions to the drag rate of processes (ai) and (aii)
  cancel each other. The drag rate is determined by the $g_{1\perp}$ and $g_{3\perp}$ processes.}
  \label{Fig:scatteringprocesses}
\end{figure*}

The scattering processes (ai) and (aii) correspond to $g_{5\perp}$ scattering in the g-ology classification,
with one of the scattering states being tied to the Dirac point. Denoting $k_1-k_{1'} = q$,
we have for the two contributions to $1/\tau_D^\infty$:
\begin{align}
\begin{split}
R_{\rm ai}=\,&-\frac{\pi}{T} \sum_{\eta_1} \int_{k_1q}\, \big|V_{12}(q,v q )\big|^2
\left(\frac{q}{k_0}\right)^4 n_F(\epsilon_1)n_F(v\eta_1q)\left\{1-n_F[v\eta_1(k_1+q)]\right\}[1-n_F(0)] \,
\end{split}\\
\begin{split}
R_{\rm aii}=\,&\frac{\pi}{T} \sum_{\eta_1} \int_{k_1k_{1'}}\, \big|V_{12}(q, v q)\big|^2
\left(\frac{q}{k_0}\right)^4 n_F(\epsilon_1)n_F(0)\left\{1-n_F[v\eta_1(k_1+q)]\right\}[1-n_F(-v\eta_1q)]~.
\end{split}
\end{align}
These are seen to cancel out, $R_{\rm ai}=-R_{\rm aii}$. This is because the initial states for
one of the two channels of $g_{5\perp}$ scattering are the final states for the other
(Fig.~\ref{Fig:scatteringprocesses}), so that the product of the thermal factors is the same for
both channels, namely
\begin{align}
n_F(\epsilon_1)n_F(v\eta_1q)\left\{1-n_F[v\eta_1(k_1+q)]\right\}[1-n_F(0)]
=n_F(\epsilon_1)n_F(0)\left\{1-n_F[v\eta_1(k_1+q)]\right\}[1-n_F(-v\eta_1q)]~,
\end{align}
whereas the transferred momenta are of opposite sign.

The high-frequency drag rate is thus determined by the remaining sum of the contributions of
the (bi) and (bii) processes, which correspond to $g_{1\perp}$ backscattering and $g_{3\perp}$
Umklapp scattering, respectively:
\begin{align}
\frac{1}{\tau^{\infty}_D} =\frac{4 \pi}{ T k_0^8}  \int_{qQ}\,
Q^4 q^4\big|V_{12}(q,2 vQ)\big|^2 \,W(q,Q)~,
\label{highfrequencydrag}
\end{align}
where
\begin{align}
W(q,Q)=\frac{1}{\cosh\frac{vq-2\mu}{2T}+\cosh\frac{vQ}{T}}
\left[\,\frac{1}{\cosh\frac{vq-2\mu}{2T}+\cosh\frac{vQ }{T}}
-\frac{1}{\cosh\frac{vq+2\mu}{2T}+\cosh\frac{vQ}{T}} \,\right] +(q\to -q)
\end{align}
\end{widetext}
and $2Q = k_1+k_{1'}$ is the total momentum of the incoming ($k_1$) and outgoing ($k_{1'}$)
particles in edge 1. The energy-momentum conservation fixes the transferred frequency in
the RPA interaction at $\epsilon_1-\epsilon_{1'} =2vQ$.

The $g_{1\perp}$ and $g_{3\perp}$ contributions to $1/\tau_D^\infty$ [which are given by
the first and second terms in the square brackets in Eq.~(\ref{highfrequencydrag}), respectively]
are of opposite sign, but---in contrast to the $g_{5\perp}$ processes, whose contribution
to the drag rate vanishes exactly for arbitrary $\mu$---they generically do not cancel each
other exactly. Note that they do so, however, at the particle-hole symmetric point $\mu=0$.

\subsection{Dynamically screened interaction}
\label{sIIIc}

Before proceeding with the calculation of the drag rate, let us elaborate on the importance
of the dynamical part of the screened RPA interaction $V_{12}(q,2 v_F Q)$ in Eq.~(\ref{highfrequencydrag}).
Derived in Appendix \ref{a1}, $V_{12}(q,\Omega)$ reads
\begin{align}
V_{12}(q,\Omega) = \frac{V_0 e^{-|q|d} \left[ (vq)^2 -\Omega^2\right]^2}
{\left[ (\Omega + i \Gamma_+)^2- \Omega_+^2 \right]
\left[ (\Omega + i \Gamma_-)^2- \Omega_-^2 \right]}~,
\label{RPAV12}
\end{align}
where the dispersion relation for the symmetric ($+$) and antisymmetric ($-$), in the edge index,
plasmon modes is given by
\begin{align}
\Omega_\pm(q) = v_{\pm}(q) |q|
\label{plasmonpolesmaintext}
\end{align}
with the velocities
\begin{align}
v_{\pm}(q) = \sqrt{1+ \alpha_{\pm}(q)} \,  v
\label{plasmonvelocitymaintext}
\end{align}
[see also Eq.~(\ref{plasmonvelocities}) for the excitation spectrum of the bosonized Hamiltonian] and
\begin{align}
\alpha_{\pm}(q) = \alpha \left(1 \pm e^{-|q| d}\right)~,
\label{plasmoninteractionmaintext}
\end{align}
with $\alpha = V_0/\pi v$ being the dimensionless strength of intrawire interactions.
It is assumed here that the Fourier components of the intraedge and interedge bare potentials
at zero momentum [$V_{0\parallel}$ and $V_{0\perp}$ below Eq.~(\ref{elementsB}), respectively]
are the same, so that the velocity of the antisymmetric plasmon at $q=0$ is not renormalized by interactions;
this makes the calculation less cumbersome without changing the result qualitatively.
The plasmon damping rate $\Gamma_\pm (q)$ is induced by chirality-changing electron scattering and,
at the lowest (second) order in $\alpha_\pm$, is written (see Appendix \ref{a1}) as
\begin{multline}
\Gamma_{\pm}(q)=\frac{1}{16}\,\alpha_{\pm}^2(q) \left(\frac{q}{k_0}\right)^4 {vq\,\sinh\frac{vq}{T}\over \cosh{vq-\mu\over 2T}\cosh{vq+\mu\over 2T}}~.
\label{plasmonratemaintext}
\end{multline}

Dynamical screening in the HLL is unusual in two key aspects.
First, the very appearance of the plasmon poles in the screened {\it backscattering} interaction [Eq.~(\ref{RPAV12})]
is a rather special feature of the HLL, which distinguishes it---quite apart from the topological protection
against elastic backscattering---from the conventional TLL model. The primary property of the HLL that is behind
this distinction is that the plasmon excitations (chiral polarization bubbles in the fermionic diagrammatic language)
are only coupled to the backscattering interaction, i.e., participate in its screening, by $g_5$ scattering
(see Appendix \ref{a1}) which, in turn, relies on the existence of a Dirac point.
That is, while the $g_{5\perp}$ processes do not, as shown in Sec.~\ref{Sec:High frequency drag},
contribute to Coulomb drag directly, they influence it by triggering the additional, plasmon-mediated mechanism
of electron-electron backscattering.

Second, the plasmon contribution to Coulomb drag is suppressed by the topological nature of the edge states
much more weakly than the electron-hole contribution. This can already be inferred from the observation
that $1/\tau_D^\infty$ in Eq.~(\ref{highfrequencydrag}), being taken with the static interaction
potential $V_{12}(q,0)$, scales with the ultraviolet momentum scale $k_0$ as $1/k_0^8$,
whereas $\Gamma_\pm$ from Eq.~(\ref{plasmonratemaintext}) scales as $1/k_0^4$.
A consequence of this is that the plasmon-mediated Coulomb drag, which is entirely due to dynamical screening,
plays a much more prominent role in the HLL compared to more conventional higher-dimensional conductors
(for more detail, see Sec.~\ref{sIIIe}).

When integrating over $Q$, the drag rate in Eq.~(\ref{highfrequencydrag}) is represented as a sum of two terms,
one coming from the the sum over the ``thermal poles" at $Q=\pm q/2+\mu/v+i\pi(2n-1)T/v$, where $n$ is an integer,
the other coming from the ``plasmon poles" of $|V_{12}(q,2vQ)|^2$. In the limit of $\Gamma_\pm (q)\ll T$, with $\Gamma_\pm(q)$
taken at the characteristic $q$ that give the main contribution to $1/\tau_D^\infty$, the two terms can be cleanly separated as,
respectively, the electron-hole ($1/\tau_{\rm eh}$) and plasmon ($1/\tau_{\rm p}$) contributions to the drag rate:
\begin{equation}
{1\over\tau_D^\infty}\to {1\over\tau_{\rm eh}}+{1\over\tau_{\rm p}}~.
\label{eh_vs_p}
\end{equation}
These two will be calculated in Secs.~\ref{sIIId} and \ref{sIIIe}.

\subsection{Electron-hole contribution to the drag rate}
\label{sIIId}

We first calculate $1/\tau_{\rm eh}$, the electron-hole contribution to the drag rate,
defined above Eq.~(\ref{eh_vs_p}) and obtainable by neglecting the dynamical part of
the screened potential in Eq.~(\ref{highfrequencydrag}). To find $1/\tau_{\rm eh}$ for $|\alpha|\ll 1$,
we substitute the bare potential $V_{12}(q)$ for the static potential $V_{12}(q,0)$.
The result depends on the relation between three energy scales, $T$, $|\mu|$, and $v/d$,
all being assumed to be much smaller than the ultraviolet scale of our model $vk_0$.

\begin{figure}
  \centering
  \includegraphics[width=.4\textwidth]{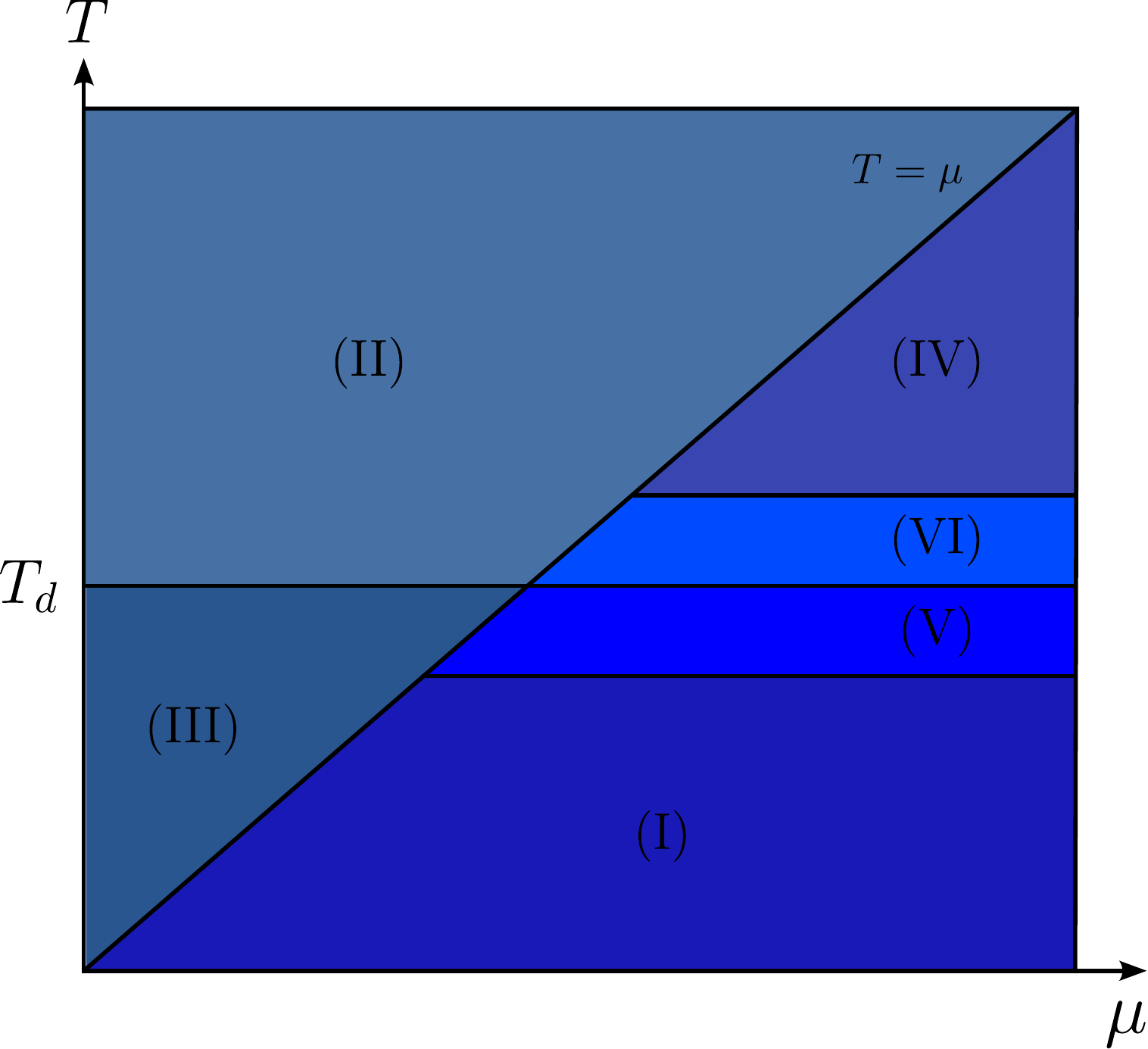}
  \caption{Electron-hole contribution $1/\tau_{\rm eh}$ to the drag rate, with different regimes
  in the $T$--$\mu$ plane labeled according to the corresponding equations in Sec.~\ref{sIIId}.
  The sequence of different types of the $T$ dependence of $1/\tau_{\rm eh}$, depending on whether
  the distance $d$ between the edges is larger or smaller than the Fermi wavelength, is shown at
  the very end of Sec.~\ref{sIIId}. In the low-$T$ limit, $1/\tau_{\rm eh}$ vanishes at $T\to 0$ as $T^5$ (regime I).
  In the high-$T$ limit, $1/\tau_{\rm eh}$ does not depend on $T$ (regime II). In between, $1/\tau_{\rm eh}$
  monotonically increases with growing $T$ at fixed $\mu$, with the crossover temperatures given by $|\mu|$ and
  (as marked on the $T$ axis) $T_d \sim v/d$. For $k_Fd\gg 1$, there is a sharp change (regimes V and VI) in the $T$
  behavior of $1/\tau_{\rm eh}$ at $T\simeq T_c=v/2d$.
    }
  \label{Fig:T-mu-plot-eh}
\end{figure}

In the limit of low $T$, for $T\ll\min\{v/d,|\mu|\}$, we obtain
\begin{multline}
{\rm (I):}\quad {1\over\tau_{\rm eh}}\simeq {64\over 5}\pi\alpha^2
\left({\mu\over vk_0}\right)^8\left({\pi T\over \mu}\right)^4T\,e^{-4k_Fd}~,\\
T\ll\min\{v/d,|\mu|\}~.
\label{d1}
\end{multline}
The main contribution to $1/\tau_{\rm eh}$ in this limit comes from $g_{1\perp}$ processes,
with the characteristic $|q\pm 2k_F|\sim |Q|\sim T/v$. The drag rate (\ref{d1}) vanishes
for $T\to 0$ as $T^5$. Here and below, (I), (II), etc. label different transport regimes
to be shown in Fig.~\ref{Fig:T-mu-plot-eh}, etc.

In the opposite limit of high $T$, for $\max\{v/d,|\mu|\}\ll T$, we get
\begin{multline}
{\rm (II):}\quad {1\over\tau_{\rm eh}}\simeq {\pi\over 5}(3\pi^4-35\pi^2+60)\\
\times\pi\alpha^2 \left({\pi T\over vk_0}\right)^8\!\left({\mu\over \pi T}\right)^2\!\left({v\over \pi Td}\right)^7T~,\\
\max\{v/d,|\mu|\}\ll T~.
\label{d2}
\end{multline}
The contributions of $g_{1\perp}$ and $g_{3\perp}$ processes to Eq.~(\ref{d2})---which are,
as already mentioned in Sec.~\ref{Sec:High frequency drag}, of different sign---strongly compensate
each other, with the characteristic $|q|\sim 1/d$ and $|Q|\sim T/v$. The structure of Eq.~(\ref{d2})
in the form of a product of four $T$ dependent factors transparently reflects the physics of
Coulomb drag in the high-$T$ limit. The $T$ dependence of $1/\tau_{\rm eh}$ that results from this product
is seen to cancel out; that is, in the limit of high $T$, the electron-hole contribution to the drag rate is
independent of $T$.

For $T$ between two other scales, $|\mu|$ and $v/d$, the result for $1/\tau_{\rm eh}$ reads, depending
on which of the two scales is larger:
\begin{multline}
{\rm (III):}\quad {1\over\tau_{\rm eh}}\simeq {512\over 315}\left({22\over 15}\pi^2+13\right)\\
\times \pi\alpha^2 \left({\pi T\over vk_0}\right)^8\left({\mu\over \pi T}\right)^2T~,\\
|\mu|\ll T\ll v/d~,
\label{d3}
\end{multline}
and
\begin{multline}
{\rm (IV):}\quad {1\over\tau_{\rm eh}}\simeq 18\pi\alpha^2 \left({\mu\over vk_0}\right)^8\!
\left({v\over Td}\right)^4\!{1\over (k_F d)^3}\,T\,e^{-{2|\mu|\over T}}~,\\
v/d\ll T\ll |\mu|~.
\label{d4}
\end{multline}
In both cases, similarly to Eq.~(\ref{d2}), there is a strong compensation between the contributions
of $g_{1\perp}$ and $g_{3\perp}$ processes. In both cases, the main contribution to $1/\tau_{\rm eh}$ comes
from $|q|\sim 1/d$: the difference is that the characteristic $|Q|$ is given by $T/v$ in Eq.~(\ref{d3})
and by $k_F$ in Eq.~(\ref{d4}). Note that $1/\tau_{\rm eh}$ in Eq.~(\ref{d4}) behaves, with changing $T$,
according to the Arrhenius law with the activation gap $2|\mu|$.

The crossover between the limits  $T\ll v/d\ll|\mu|$ and $v/d\ll T\ll|\mu|$ [Eqs.~(\ref{d1}) and (\ref{d4}),
respectively] has the form of a sharp singularity at $T=T_c$, where $T_c=v/2d$. Specifically:
\begin{multline}
{\rm (V):}\quad {1\over\tau_{\rm eh}}\simeq 192\pi\alpha^2\left({\mu\over vk_0}\right)^8\left({T_c\over \mu}\right)^4T_c\\
\times e^{-{2|\mu|\over T_c}}\left({T_c\over T_c-T}\right)^6~,\\ {v\over k_Fd^2}\ll T_c-T\ll T_c
\label{d5}
\end{multline}
for $T<T_c$ and
\begin{multline}
{\rm (VI):}\quad {1\over\tau_{\rm eh}}\simeq {384\over 5}\pi\alpha^2\left({\mu\over vk_0}\right)^8\left({T_c\over |\mu|}\right)^3T_c\\
\times e^{-{2|\mu|\over T}}\left({T_c\over T-T_c}\right)^5~,\\
{v\over k_Fd^2}\ll T-T_c\ll T_c
\label{d6}
\end{multline}
for $T>T_c$, both in the ``critical region" $|T-T_c|\ll T_c$. The broadening of the
power-law ``resonance" at $T=T_c$ is of the order of $v/k_Fd^2\ll T_c$.

The peculiar behavior of $1/\tau_{\rm eh}$ in Eqs.~(\ref{d5}) and (\ref{d6}) is related to the interplay
between two exponential factors in the integrand of Eq.~(\ref{highfrequencydrag}): $\exp (-2|q|d)$
[from the static interaction potential $V_{12}(q)$] and $\exp (-v|q\pm 2k_F|/2T)$
(from the thermal occupation factors for $T\ll |\mu|$) in the interval $-2k_F<q<2k_F$.
On the low-$T$ [Eq.~(\ref{d5})] and high-$T$ [Eq.~(\ref{d6})] sides of the resonance,
the integrand is sharply peaked at $|q|=2k_F$ and $q=0$, respectively.
Specifically, for Eq.~(\ref{d5}), the characteristic momenta are $|q\pm 2k_F|\sim |Q|\sim T_c^2/v(T_c-T)$.
For Eq.~(\ref{d6}), they are $|q|\sim T_c^2/v(T-T_c)$ and $|Q|\sim k_F$. At the resonance,
the $q$ dependence in the exponential factors cancels out and $1/\tau_{\rm eh}$ is determined
by $|q|\sim |Q|\sim k_F$. It is worth noting that, despite Eqs.~(\ref{d5}) and (\ref{d6})
having the spike-like power-law factors, $1/\tau_{\rm eh}$ is a monotonic function of $T$,
namely it increases with growing $T$ on both sides of the ``spike." This is because the
nonsingular (at $T=T_c$) factor $\exp (-2|\mu|/T)$ in Eq.~(\ref{d6}) is a faster function of $T$
than the singular factor $[T_c/(T-T_c)]^5$ in the tail of the resonance for $T-T_c\gg v/k_Fd^2$.

As $T$ increases, the sequence of different types of the $T$ dependence of $1/\tau_{\rm eh}$---depending
on whether $k_Fd\gg 1$ or $k_Fd\ll 1$---is as follows (see also Fig.~\ref{Fig:T-mu-plot-eh}).
For $k_Fd\gg 1$, $1/\tau_{\rm eh}$ behaves as
\begin{multline*}
T^5 \,\,{\rm (I)} \,\to \,{\rm sharp\,\, crossover\,\, (V)\!+\!(VI)}\\
\to \,\exp(-2|\mu|/T)/T^3 \,\,{\rm  (IV)}\, \to \,{\rm const}(T)\,\, {\rm (II)}~.
\end{multline*}
For $k_Fd\ll 1$, $1/\tau_{\rm eh}$ behaves as
\begin{align*}
T^5 \,\,{\rm (I)} \,\to \,T^7\,\, {\rm (III)} \,\to \,{\rm const}(T) \,\,{\rm  (II)}~.
\end{align*}
In both cases, $1/\tau_{\rm eh}$ is a monotonic function of $T$, vanishing at $T\to 0$
and saturating in the limit of large $T$.

\subsection{Plasmon-mediated Coulomb drag}
\label{sIIIe}

We now turn to the calculation of $1/\tau_{\rm p}$, the plasmon contribution to the drag
rate $1/\tau_D^\infty$, as defined above Eq.~(\ref{eh_vs_p}). Taking the residue of the four plasmon
poles in the lower half-plane of $Q$ at $Q=(v_\pm/2v)q-i\Gamma_\pm/2v$ and $Q=-(v_\pm/2v)q-i\Gamma_\pm/2v$,
we represent $1/\tau_{\rm p}$ for $|\alpha|\ll 1$ as
\begin{equation}
\begin{split}
&{1\over\tau_{\rm p}}\simeq {\pi^3\over 128}\,\alpha^2{v^5\over Tk_0^8}\int_q \,q^{12}e^{-2|q|d}\,W(q,q/2)\\
&\times\sum_\pm \left(1-{v_\pm^2\over v^2}\right)^4\!{1\over \Gamma_\pm}\,
{\rm Re}\,{1\over \left[(v_+-v_-)q-i\Gamma_\pm\right]^2+\Gamma_\mp^2}~.
\end{split}
\label{e1}
\end{equation}
In the derivation of Eq.~(\ref{e1}), we assumed that the plasmon modes are weakly decaying.
More precisely, we used not only the condition $v_\pm|q|\gg \Gamma_\pm$ which is altogether
necessary to meaningfully define the plasmon modes in the first place, but a stronger
condition $|\alpha_\pm v_\pm q|\gg \Gamma_\pm$. The latter makes it justifiable to neglect
the plasmon damping in the factor $q^2-4Q^2$ in the numerator of $V_{12}(q,2Q/v)$ [Eq.~(\ref{RPAV12})]
at the plasmon poles in Eq.~(\ref{highfrequencydrag}), not requiring at the same time
that $|(v_+-v_-)q|$ is large compared to $\Gamma_\pm$.

The important point here is that the energy splitting between the symmetric and antisymmetric
plasmon modes $|(v_+-v_-)q|$ falls off sharply with increasing $|q|d$, namely as $\exp (-2|q|d)$,
so that the broadening of the modes $\Gamma_\pm$---even though being small in the
parameter $(q/k_0)^4\ll 1$ and the additional power of $\alpha$ [Eq.~(\ref{plasmonratemaintext})]---can
become larger than the splitting for $|q|$ larger than a certain characteristic momentum $q_p\gg 1/d$
(with $q_p$ being still much smaller than $k_0$). As will be shown below, this circumstance essentially
modifies the general picture of plasmon-mediated Coulomb drag, with friction being strongly suppressed
by the overlap between the symmetric and antisymmetric plasmon modes.
For $(\alpha_+-\alpha_-)^2v^2q^2\gg |\Gamma_+^2-\Gamma_-^2|$ (which includes $|q|\sim q_p$),
the second line in Eq.~(\ref{e1}) can be further simplified to
\begin{align}
\sum_\pm {4\alpha_\pm^4\over \Gamma_\pm}\,{1\over (\alpha_+-\alpha_-)^2q^2+16\Gamma_\pm^2}~.
\label{e2}
\end{align}

Let us first calculate $1/\tau_{\rm p}$ for the case in which the main contribution to $1/\tau_{\rm p}$
comes from $|q|\ll q_p$, i.e., the plasmon damping can be neglected in the ``Lorentzian" (\ref{e2})
(recall that $\Gamma_\pm$ is a strong function of $q$, hence the quotation marks). Importantly, the
exponential factor $e^{-2|q|d}$ in Eq.~(\ref{e1}) is then canceled by the same factor in $(\alpha_+-\alpha_-)^2$
from the denominator in Eq.~(\ref{e2}), so that the screened interaction effectively extends beyond the
scale of $1/d$ in $q$ space. Specifically, Eq.~(\ref{e1}) reduces to
\begin{multline}
{1\over\tau_{\rm p}}\simeq {\pi^3\over 32}\,\alpha^2{v^2\over Tk_0^4}\,{\sinh {\mu\over T}\over\cosh^3\!{\mu\over 2T}}
\int_q \,q^5\left(1+e^{-2|q|d}\right)\\
\times {1\over \cosh {vq\over 2T}}\,\left[{1\over \cosh{vq-\mu\over 2T}}-(q\to -q)\right]~.
\label{e3}
\end{multline}
The radius of the bare interaction only remains in the factor $1+\exp(-2|q|d)$ that originates from
the sum $\alpha_+^4+\alpha_-^4$ and changes between 2 and 1 with increasing $|q|$. Crucially,
the strength of interaction cancels out in Eq.~(\ref{e2}), so that $1/\tau_{\rm p}$ in Eq.~(\ref{e3})
scales with $\alpha$ as $\alpha^2$, similar to $1/\tau_{\rm eh}$.

In the limit of low $T$, for $T\ll |\mu|\ll v/d$, we obtain
\begin{multline}
{\rm (VII):}\quad
{1\over\tau_{\rm p}}\simeq {\pi^3\over 6}\pi\alpha^2\left({\mu\over vk_0}\right)^4
\left({\mu\over\pi T}\right)^2T\,e^{-|\mu|/T}~,\\
T\ll |\mu|\ll v/d~.
\label{e4}
\end{multline}
In stark contrast to the electron-hole contribution to the drag rate, which
vanishes for $T\to 0$ as a power-law of $T$, the plasmon contribution in Eq.~(\ref{e4})
obeys Arrhenius' law. The main contribution to Eq.~(\ref{e4}) comes from all $q$ in the
interval $0<|q|<k_F$.

For $|\mu|\ll T\ll v/d$, we have
\begin{multline}
{\rm (VIII):}\quad
{1\over\tau_{\rm p}}\simeq {7\pi^3\over 24}\pi\alpha^2\left({\pi T\over vk_0}\right)^4
\left({\mu\over \pi T}\right)^2T~,\\
|\mu|\ll T\ll v/d~,
\label{e5}
\end{multline}
with characteristic $|q|\sim T/v$. For $v/d\ll\max\{T,|\mu|\}$, the exponential
term $e^{-2|q|d}$ in Eq.~(\ref{e3}) can be neglected, so that $1/\tau_{\rm p}$ is obtained
by multiplying Eq.~(\ref{e4}) (for $T\ll v/d\ll|\mu|$) or Eq.~(\ref{e5}) (for $|\mu|\ll v/d\ll T$)
by a factor of 1/2.

Now, turn to the case of $\max\{T,|\mu|\}\gg T_p=vq_p/2$, where the plasmon damping
substantially modifies plasmon-mediated Coulomb drag. In this limit, $1/\tau_{\rm p}$
can be represented as
\begin{multline}
{1\over\tau_{\rm p}}\simeq {\pi^3\over 32}\,\alpha^2{v^2\over Tk_0^4}\,
{\sinh {\mu\over T}\over\cosh^3\!{\mu\over 2T}}\int_q \,q^5 {1\over s(q)}\\
\times {1\over \cosh {vq\over 2T}}\,\left[{1\over \cosh{vq-\mu\over 2T}}-(q\to -q)\right]~,
\label{e6}
\end{multline}
where
\begin{multline}
s(q)=1+4\alpha^2\left({2q\over k_0}\right)^8e^{2|q|d}\\
\times {\sinh^2\! {vq\over T}\over \cosh^2{vq-\mu\over 2T}\,\cosh^2{vq+\mu\over 2T}}~.
\label{e7}
\end{multline}
In fact, Eq.~(\ref{e6}) has a broader range of applicability, namely $\max\{T,|\mu|\}\gg v/d$.
In particular, it gives, for $s(q)\to 1$, the result that follows from Eq.~(\ref{e3})
for $v/d\ll\max\{T,|\mu|\}\ll T_p$, as discussed below Eq.~(\ref{e5}).
What changes for $\max\{T,|\mu|\}\gg T_p$ is that the function $s(q)$ can no longer be approximated
by unity for $|q|\sim \max\{T/v,k_F\}$. Specifically, $1/s(q)$ behaves as a step function:
$1/s(q)\to\theta (q_p-|q|)$, falling off sharply with increasing $|q|$, as $e^{-2|q|d}$,
for $|q|-q_p\gg 1/d$, where $q_p\gg 1/d$ is defined by $s(q_p)-1\sim 1$.
That is, $1/\tau_{\rm p}$ in the limit of strong plasmon damping is determined by all $q$
in the interval $0<|q|<q_p$, with $q_p\ll\max\{T/v,k_F\}$.

We obtain, for three different regimes of plasmon-mediated Coulomb drag in which the plasmon
damping is important:
\begin{multline}
{\rm (IX):}\quad
{1\over\tau_{\rm p}}\simeq {2\over 7\pi}\pi\alpha^2\left({\pi T\over vk_0}\right)^4
\left({\mu\over \pi T}\right)^2\left({T_p\over T}\right)^7T~,\\
\max\{|\mu|,T_p\}\ll T~,
\label{e8}
\end{multline}
where
\begin{align}
T_p={v\over 2d}\ln\left[\,{(k_0d)^4\over |\alpha|}\,{Td\over v}\,\right]~;
\label{e9}
\end{align}
\begin{multline}
{\rm (X):}\quad
{1\over\tau_{\rm p}}\simeq {16\over 3\pi^3}\pi\alpha^2\left({\pi T\over vk_0}\right)^4
\left({T_p\over T}\right)^6T\,e^{-|\mu|/T}~,\\
T\ll T_p\ll |\mu|~,
\label{e10}
\end{multline}
where
\begin{align}
T_p={v\over 2d}\left\{{|\mu|\over T}+\ln\left[\,{(k_0d)^4\over |\alpha|}\,\right]\right\}~;
\label{e11}
\end{align}
and
\begin{multline}
{\rm (XI):}\quad
{1\over\tau_{\rm p}}\simeq {32\over 7\pi^3}\pi\alpha^2\left({\pi T\over vk_0}\right)^4
\left({T_p\over T}\right)^7T\,e^{-|\mu|/T}~,\\
T_p\ll T\ll |\mu|~,
\label{e12}
\end{multline}
where
\begin{align}
T_p={v\over 2d}\left\{{|\mu|\over T}+\ln\left[\,{(k_0d)^4\over |\alpha|}\,{Td\over v}\,\right]\right\}~.
\label{e13}
\end{align}

The term $|\mu|/T$ in Eqs.~(\ref{e11}) and (\ref{e13}) for $T_p$ appears because, for $T\ll |\mu|$,
the plasmon damping rate obeys the Arrhenius law with the activation gap $|\mu|$. Specifically:
\begin{align}
\Gamma_\pm(q)\simeq {1\over 4}\,\alpha^2_\pm(q)\,T\left({vq\over T}\right)^2\!
\left({q\over k_0}\right)^4\!e^{-|\mu|/T}
\label{e14}
\end{align}
for $v|q|\ll T$, which is the condition relevant to Eq.~(\ref{e12}), and
\begin{align}
\Gamma_\pm(q)\simeq {1\over 8}\,\alpha^2_\pm(q)\,v|q|\left({q\over k_0}\right)^4\!
\exp \left({v|q|-|\mu|\over T}\right)
\label{e15}
\end{align}
for $T\ll v|q|<|\mu|$ (more precisely, as far as the right condition is concerned,
for $|\mu|-v|q|\gg T$), which is the condition relevant to Eq.~(\ref{e10}).

The plasmon damping is seen to strongly suppress plasmon-mediated Coulomb drag when
the energy splitting between the symmetric and antisymmetric plasmon modes
(for $|q|\sim \max\{T/v,k_F\}$) becomes much smaller than their damping rate.
Specifically, $1/\tau_{\rm p}$ for $|\mu|\ll T$ is suppressed in Eq.~(\ref{e8})
compared to Eq.~(\ref{e5}) by the additional factor $(T_p/T)^7\ll 1$. For $T\ll |\mu|$,
the suppression factor, compared to Eq.~(\ref{e4}), is $(T_p/|\mu|)^6\ll 1$ in Eq.~(\ref{e10})
and $(T_p/|\mu|)^6T_p/T\ll 1$ in Eq.~(\ref{e12}). In all the cases, $1/\tau_{\rm p}$
vanishes as a power law of $T_p$ with increasing damping rate.

\begin{figure}
  \centering
  \includegraphics[width=.4\textwidth]{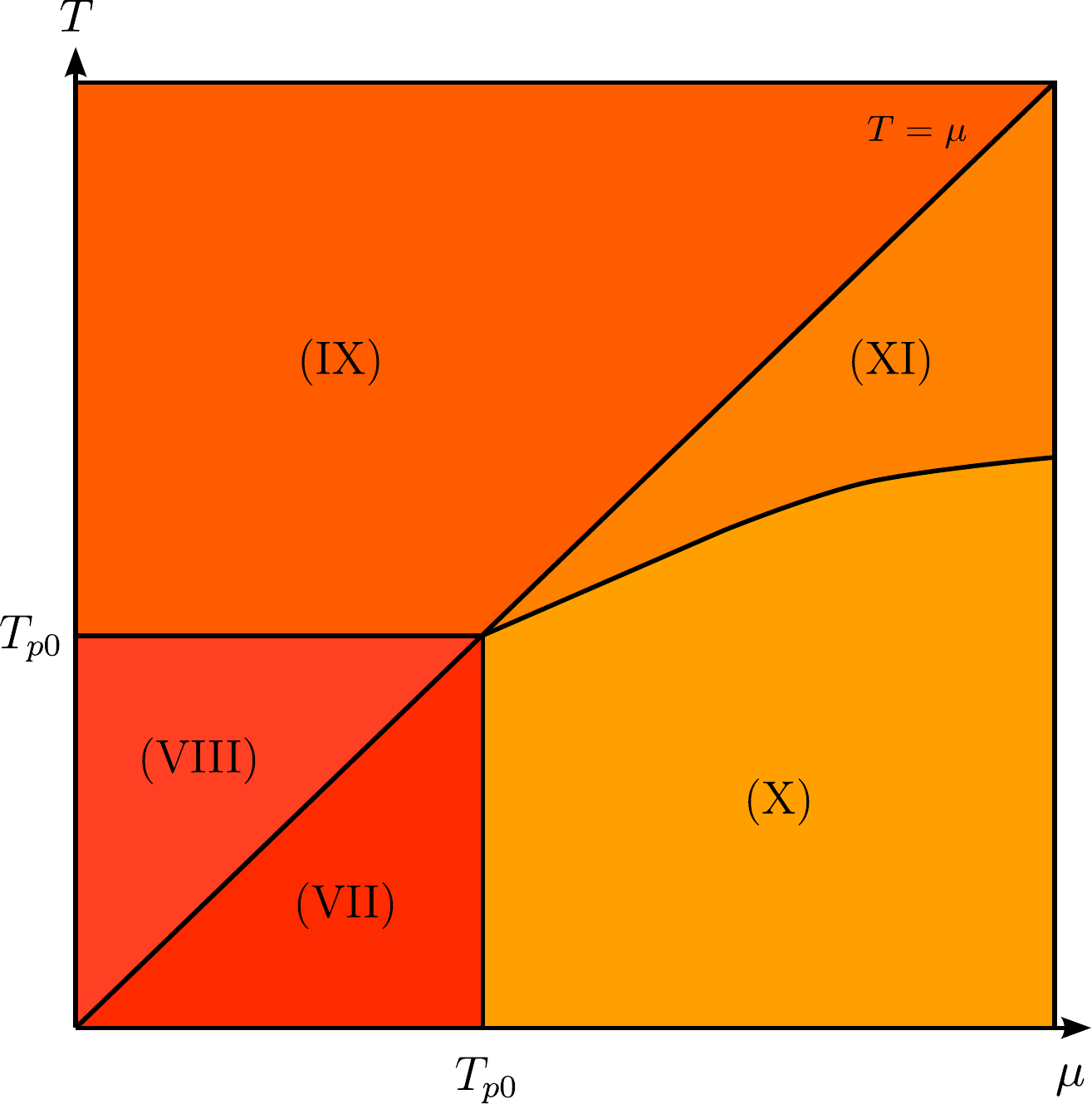}
  \caption{Plasmon contribution $1/\tau_{\rm p}$ to the drag rate, with different regimes
  in the $T$--$\mu$ plane labeled according to the corresponding equations in Sec.~\ref{sIIIe}.
  The sequence of different types of the $T$ dependence of $1/\tau_{\rm p}$, depending on the
  relation between $|\mu|$ and the characteristic temperature $T_{p0}$ [Eq.~(\ref{e16})],
  is shown at the end of Sec.~\ref{sIIIe}. In contrast to the electron-hole contribution to
  the drag rate (Fig.~\ref{Fig:T-mu-plot-eh}), $1/\tau_{\rm p}$ is a nonmonotonic function
  of $T$ for given $\mu$. In the low-$T$ limit, $1/\tau_{\rm p}$ vanishes at $T\to 0$
  according to Arrhenius' law with the activation energy $|\mu|$ (regimes VII and X).
  In the high-$T$ limit, $1/\tau_{\rm p}$ vanishes with increasing $T$ as $T^{-4}$ (regime IX).
    }
  \label{Fig:T-mu-plot-pl}
\end{figure}

Recall that the energy $T_p$, being defined in terms of the momentum scale $q_p$
above which the plasmon damping becomes relevant, is a function of $T$.
This means that the crossover temperatures that separate between regimes VIII and IX on the one hand
and between regimes X and XI on the other follow as the solution of the equation $T_p(T)=T$.
One of the characteristic scales of $T$ that emerges from this is
\begin{align}
T_{p0}={v\over 2d}\ln {(k_0d)^4\over |\alpha|}~.
\label{e16}
\end{align}
The other is
\begin{align}
T_{p1}={T_{p0}\over 2}\left(1+\sqrt{1+{v\over 2d}\,{|\mu|\over T_{p0}^2}}\right)~.
\label{51a}
\end{align}
Depending on the relation between $|\mu|$ and $T_{p0}$, there are two distinct sequences,
with varying $T$, of different types of the $T$ dependence of $1/\tau_{\rm p}$
(see also Fig.~\ref{Fig:T-mu-plot-pl}). For $|\mu|\ll T_{p0}$, as $T$ is increased,
$1/\tau_{\rm p}$ first increases according to the Arrhenius law (VII) and then keeps
growing, as $T^3$ (VIII), before the growth changes to the $1/T^4$ (IX) falloff:
\begin{align*}
e^{-|\mu|/T}/T\,\,{\rm (VII)}\,\to\, T^3\,\,{\rm (VIII)}\,\to\, T^{-4}\,\, {\rm (IX)}~.
\end{align*}
For $T_{p0}\ll |\mu|$, the interval of $T$ within which there was the $T^3$ behavior
of $1/\tau_{\rm p}$ in the opposite limit shrinks to zero. The activation growth
of $1/\tau_{\rm p}$ with increasing $T$ [(X) and (XI), with different preexponential
factors in the Arrhenius law] is directly followed by the downturn to the $1/T^4$ (IX) behavior:
\begin{multline*}
e^{-|\mu|/T}T_p^6(T)/T\,\,{\rm (X)}\,\to\, e^{-|\mu|/T}T_p^7(T)/T^2\,\,{\rm (XI)}\\
\to\, T^{-4}\,\, {\rm (IX)}~.
\end{multline*}
As can be seen from Fig.~\ref{Fig:T-mu-plot-pl}, the energy $T_{p0}$ also gives
the crossover scale for $|\mu|$ when it varies between regimes VII and X.
The temperature $T_{p1}$ as a function of $\mu$ [Eq.~(\ref{51a})] gives the boundary
between regimes X and XI.

By comparing the results for $1/\tau_{\rm eh}$ (I-VI) on the one hand and $1/\tau_{\rm p}$ (VII-XI) on the other,
the most notable differences between the electron-hole and plasmon contributions to the drag rate are the following.
First of all, as already noted in Sec.~\ref{sIIIc}, the two are different in the way they scale
with the ultraviolet momentum cutoff of our theory,
namely $1/\tau_{\rm eh}\propto 1/k_0^8$ and $1/\tau_{\rm p}\propto 1/k_0^4$.
This renders $1/\tau_{\rm eh}$ to be much smaller than $1/\tau_{\rm p}$
when the three characteristic energy scales $T$, $|\mu|$, and $v/d$ are of the same order.
Moreover, one can see that $1/\tau_{\rm eh}\ll 1/\tau_{\rm p}$ at $T\sim |\mu|$ for arbitrary $k_Fd$.
This brings us to the question of differences in the $T$ dependence of $1/\tau_{\rm eh}$ and $1/\tau_{\rm p}$.

One of the differences is that $1/\tau_{\rm p}$ is a nonmonotonic function of $T$, i.e., in the high-$T$ limit, $1/\tau_{\rm eh}$
is independent of $T$ (II), whereas $1/\tau_{\rm p}$ decreases with increasing $T$ (IX). However, on the side of high $T$,
the plasmon-mediated mechanism of Coulomb drag can be seen to remain dominant, with $1/\tau_{\rm p}\gg 1/\tau_{\rm eh}$
in the whole range of $T$ up to $T\sim vk_0$. The situation is different in the low-$T$ limit. Here, the $T$ dependence of $1/\tau_{\rm p}$
is characterized by the activation gap $|\mu|$ [(VII) and (X)], whereas $1/\tau_{\rm p}$ vanishes at $T\to 0$ as a power law of $T$.
In fact,  $1/\tau_{\rm eh}$ also behaves, similarly to $1/\tau_{\rm p}$, according to Arrhenius' law for $T\ll |\mu|$---even with the doubled
activation gap $2|\mu|$---if $k_Fd\gg 1$, but only within the intermediate interval of $T$ (IV). That is, the ``electron-hole mechanism"
of Coulomb drag inevitably wins over the plasmon mechanism in the limit of low $T$, leading to the universal $T^5$ behavior (I) of the
drag rate at $T\to 0$ for arbitrary $k_Fd$. One of the conclusions that follow from this comparison is that there necessarily exists a
crossover temperature $T_1$ which separates the electron-hole (lower $T$) and plasmon-dominated (higher $T$) regimes of Coulomb drag. Specifically,
\begin{align}
T_1={1\over 4}\,{|\mu|\over \ln (k_0/|\mu|)+k_Fd}~.
\label{51}
\end{align}
Note that the saturation of the dependence of $T_1$ with increasing $|\mu|$ occurs at $|\mu|\simeq (v/d)\ln (k_0d)$, i.e., below $T_{p0}$ [Eq.~(\ref{e16})].
A similar crossover of the drag resistivity between the particle-hole dominated and plasmon-dominated regimes was predicted
in the context of two-dimensional heterostructure bilayers of strongly correlated electron liquids, within the Boltzmann-Langevin
stochastic kinetic equation approach, in Ref.~\cite{Chen_2015}.

\subsection{DC Coulomb drag}
\label{Sec: Low frequency drag}

In Secs.~\ref{sIIId} and \ref{sIIIe}, we calculated the drag rate in the high-frequency limit $1/\tau_D^\infty$.
Now we turn to Coulomb drag in the dc limit, characterized by the dc drag rate $1/\tau_D$ (related to the dc drag
resistivity $\rho_D$ by $\rho_D=\pi/e^2v\tau_D$). Generically, the relaxation rate need not be a constant of $\omega$
(when this is the case, at the model level, the system is said to obey the Drude law). In fact, the drag rate is known
to be sensitive to the rate of thermal equilibration inside each of the two conductors---to the extent that,
while being finite in the high-$\omega$ limit, the drag rate may exactly vanish at $\omega\to 0$ if some of the
thermalization processes are quenched \cite{Dmitriev_Gornyi_Polyakov_2012,Narozhny_Levchenko_2015}
(see also Refs.~\cite{Schuett_2011,Narozhny_2012,dmitriev16} for other examples of a failure of the perturbative
approach to Coulomb drag). By the same token, the Drude law {\it is} valid for Coulomb drag if the thermalization
rate inside each of the conductors is much larger than the drag rate \cite{Dmitriev_Gornyi_Polyakov_2012}.

In the HLL, the intraedge electron-electron scattering rate $1/\tau_{ee}$, resulting from $g_{5\parallel}$
interactions, reads  \cite{Kainaris_2014}
\begin{align}
   \frac{1}{\tau_{ee}} \sim  \alpha^2  \left( \frac{ T}{ v_F k_0}\right)^4  T \, , \qquad  |\mu|\alt T
\label{f1}
\end{align}
for $|\mu|\alt T$ and
\begin{align}
   \frac{1}{\tau_{ee}} \sim  \alpha^2  \left( \frac{ \mu}{  vk_0}\right)^4  |\mu| \; e^{-|\mu|/T } \, , \qquad  T \alt |\mu|
   \label{dcrelaxation}
\end{align}
for $T\alt |\mu|$. This is the thermalization rate that should be compared with $1/\tau_D^\infty$ obtained in
Secs.~\ref{sIIId} and \ref{sIIIe}. For $|\mu|\ll T$, the drag rate is mainly given by $1/\tau_{\rm p}$,
which is smaller than $1/\tau_{ee}$ from Eq.~(\ref{f1}) by a small factor $(\mu/T)^2$ for $T\ll T_p$ [Eq.~(\ref{e5})]
and is further suppressed by plasmon damping [Eq.~(\ref{e8})]. It follows that the dc drag rate $1/\tau_D$ for $|\mu|\ll T$
coincides with $1/\tau_{\rm p}$ calculated in Sec.~\ref{sIIIe}.

For $T\ll |\mu|$, both $1/\tau_{\rm p}$ and $1/\tau_{ee}$ obey Arrhenius' law with the same activation gap $|\mu|$;
however, the pre-exponential factors behave with varying $T$ differently. If $|\mu|\ll T_p$, then $1/\tau_{\rm p}\gg 1/\tau_{ee}$
for all $T\ll |\mu|$. Otherwise, the range of $T$ within which $1/\tau_{\rm p}\ll 1/\tau_{ee}$ extends,
as follows from Eq.~(\ref{e10}), down below $|\mu|$ to $T\gg |\mu|(T_p/\mu)^6$, but $1/\tau_{\rm p}$ still becomes
larger than $1/\tau_{ee}$ for lower $T$. Moreover, in contrast to both $1/\tau_{ee}$ and $1/\tau_{\rm p}$,
the drag rate $1/\tau_{\rm eh}$ behaves in the low-$T$ limit as a power law of $T$ [Eq.~(\ref{d1})].
As a result, although $1/\tau_{\rm eh}\ll 1/\tau_{ee}$ at $T\sim |\mu|$ (independently of the parameter $k_Fd$),
there exists a crossover temperature---much smaller than $|\mu|$---below which the relation between the two scattering
rates is reversed. In fact, with logarithmic accuracy, this crossover temperature is given by Eq.~(\ref{51}).
Thus, irrespective of the relation between $1/\tau_{ee}$ and $1/\tau_{\rm p}$ at $T\sim |\mu|$, the drag rate is
larger than the intraedge thermalization rate in the low $T$ limit. It follows that $1/\tau_D^\infty$ and $1/\tau_D$
need not coincide in this limit.

In fact, in one dimension, the relation between $1/\tau_D^\infty$ and $1/\tau_D$ is largely dictated by the
relative strength of backscattering compared to scattering with small momentum transfer. If the latter is dominant,
then $1/\tau_D^\infty$ and $1/\tau_D$ can be vastly different from each other, as shown
in Ref.~\cite{Dmitriev_Gornyi_Polyakov_2012} for the case of one-dimensional electrons with a parabolic dispersion relation.
Otherwise, $1/\tau_D^\infty$ and $1/\tau_D$ are generically of the same order of magnitude, being only different by a numerical
coefficient. Below, we demonstrate this by solving the kinetic equation for the HLL in the low-$T$ limit, namely $T\ll v/d\ll |\mu|$.
Recall that Umklapp ($g_{3\perp}$) scattering is strongly suppressed for $T\ll |\mu|$ (Sec.~\ref{sIIId}), so that the only
scattering channel that remains (and will only be present in the collision integral below) is $g_{1\perp}$ backscattering.
It is also worth noting that forward ($g_{4\perp}$) scattering between chiral electrons is exactly absent in the collision
integral for the HLL model with a linear electron spectrum. This is because of the RPA resummation that splits the electron
and plasmon velocities in the energy-momentum conservation law and regularizes to zero the collision integral, similarly to
the spinful TLL model \cite{Yashenkin_2008}.

The collision integral (\ref{CI2}) for $g_{1\perp}$ backscattering ($\eta_1=-\eta_{1'}=-\eta_2=\eta_{2'}$) reads
\begin{multline}
     \text{st}_{1}(1) =\frac{1}{8 vk_0^8} \int_{k_{1'}} \, V_{12}^2(k_1-k_{1'})  (k_1^2-k_{1'}^2)^4\zeta^2(-\eta_1,k_{1'})
   \\ \times   \left[ g_1(-\eta_1,k_{1'}) + g_2(\eta_1,k_1) \right.\\ \left.
   -  g_1(\eta_1,k_1) - g_2(-\eta_1,k_{1'}) \right]
\end{multline}
and $\text{st}_2(1)=-\text{st}_1(1)$, at the lowest order in the static interaction (for $|\alpha|\ll 1$ and $T\ll T_1$).
The contribution of intraedge ($g_{5\parallel}$) scattering to $\text{st}_{1}(1)$ is neglected in the low-$T$ limit (for $T\ll T_1$).
By introducing the functions
\begin{align}
g_{\pm}(\eta,k)={1\over 2}\left[\,g_1(\eta,k)\pm g_{2}(\eta,k)\,\right]~,
\end{align}
we define the total and relative charge components of the distribution function
\begin{align}
g_{\pm}^c(k)={1\over 2}\left[\,g_{\pm}(+,k) - g_{\pm}(-,-k)\,\right]~.
\label{spincharge}
\end{align}
The drag resistivity can be written as \cite{Dmitriev_Gornyi_Polyakov_2012}
\begin{align}
   \rho_D = \frac{E_1-E_2}{j_-}~,
   \label{rD}
\end{align}
where the relative current $j_- = (j_1-j_2)/2$ is expressed solely in terms of $g_-^c(k)$:
\begin{align}
   j_- =  \frac{e v}{2} \int_k \, \frac{g_-^c(k)}{\cosh^2\frac{v k-\mu}{2 T}}~.
   \label{j-}
\end{align}

The kinetic equation for $g_-^{c}$ reads
\begin{multline}
   -i \omega g_-^c(k_1) - \frac{e (E_1-E_2) v}{2 T}   = \; \text{st}_c(1) \, , \label{KE4}\\
      \text{st}_c \simeq - \frac{1}{4 v k_0^8} \int_{k_{1'}}
    V_{12}^2(k_1-k_{1'}) \,\frac{(k_1^2-k_{1'}^2)^4}{\cosh^2 \frac{v k_{1'}+\mu}{2 T} } \\
                            \times
   \left[ g_-^c(-k_{1'}) + g_-^c(k_{1})  \right] \, .
\end{multline}
For $T \ll |\mu|$, we can replace $k_1-k_{1'}$ with $2k_F$
in the factors $(k_1^2-k_{1'}^2)^4$ in the collision integral.
Furthermore, for $T\ll v/d$ the argument of the interaction potential can also be set equal to $2k_F$.
Note that both these conditions on temperature are satisfied for $T\ll T_1$.
Taking the limit $\omega \to 0$, we cast
Eq.~(\ref{KE4}) in the form of a dimensionless integral equation
\begin{align}
   \mathcal{A}(x)\mathcal{G}(x) &=  1- \int \! \mathrm{d} y \, \frac{(x-y)^4}{\cosh^2 y}
\mathcal{G}(y)  \, ,\\
   \mathcal{A}(x) &= \int \! \mathrm{d} y \, \frac{(x-y)^4}{\cosh^2 y} = \frac{7 \pi^4}{120} + \pi^2 x^2 + 2 x^4 \, ,
   \label{integralequation2}
\end{align}
for the function
\begin{multline}
   \mathcal{G}(x) = \frac{2^7}{\pi}\left( \frac{k_F}{k_0}\right)^4 \frac{V_{12}^2(2 k_F)}{v^2} \,\frac{T^2}{e v (E_1-E_2)} \\
\times
\left( \frac{ T}{v k_0}\right)^4  g_-^c\!\!\left( \frac{2T }{v}x + k_F \right) \, .
\label{G}
\end{multline}
The dc drag resistivity is then obtained as
\begin{align}
   \rho_D = \frac{2}{e^2\lambda} \frac{V_{12}^2(2 k_F)}{v^2} \left(\frac{2k_F}{k_0}\right)^4 \left(\frac{2T}{v k_0}\right)^4 T \, ,
   \label{dcdrag1}
\end{align}
with the constant
\begin{align}
\lambda = \int \! \mathrm{d} x \, \frac{\mathcal{G}(x)}{\cosh^2 x} \simeq  0.242  \, .
\end{align}
We thus conclude that, for $T\alt T_1$, the drag rate that determines the dc drag resistivity
is parametrically the same as $1/\tau_D^\infty$.

\section{Intraedge interaction: Bosonization framework}
\label{LL-boson}

In Sec.~\ref{Sec:Coulomb drag in helical edges}, we discussed Coulomb drag between helical edges
within the kinetic equation framework and
neglected the TLL renormalization effects. Below, we complement the formalism of
Sec.~\ref{Sec:Coulomb drag in helical edges} by employing the bosonization
approach. In particular, this allows us to proceed to lower temperatures for which the renormalization
leads, as is usual in one dimension, to anomalous power-law $T$ dependences of the observables.

Since the renormalization effects in Luttinger liquids necessarily involve backscattering processes,
the characteristic temperature scale at which the renormalization starts cannot exceed $v/d$.
Indeed, the distance $d$ between the edges gives the characteristic radius of the interedge
interaction potential $V_{12}$, so that on smaller spatial scales one cannot treat this interaction as local.
In fact, in addition to $d$, screening by external gates introduces another spatial scale $d_0$ for both the
intraedge and interedge interaction, so that the renormalization is only operative for $T\alt v/\text{max}\{d,d_0\}$.
Below, for simplicity, we assume that $d\sim d_0$.

In what follows, we first bosonize the model and analyze the resulting phase diagram for two
coupled helical edges. Next, we discuss the implications of the renormalization effects for
the drag resistivity.

\subsection{First-order backscattering}
\label{s4}

For concreteness, we concentrate on the case of
$k_F d\gg 1$. Then, for $T\ll v/d$ (which is, as mentioned above, the range of $T$ where the renormalization
is effective) we have also $T\ll |\mu|$, so that the transitions resulting
from Umklapp interactions of two particles in the vicinity of the Dirac point are thermally suppressed.
Neglecting them, the part of Eq.~(\ref{model}) that describes chirality-changing interactions
reduces to backscattering in the vicinity of the Fermi surface.
The Hamiltonian density simplifies, then, to $\mathcal{H} = \mathcal{H}_0+ \mathcal{H}_f +\mathcal{H}_b$,
where $\mathcal{H}_0$ corresponds to the free Hamiltonian in Eq.~(\ref{kineticpart}) and $\mathcal{H}_{f,b}$
describes forward $(f)$ and backward $(b)$ scattering. Moreover, one can describe this type of backscattering
by momentum-independent coupling constants determined by the Fourier transform of the interedge and intraedge
interaction potentials at the transferred momentum
equal to $2k_F$.

To write $\mathcal{H}$, it is convenient to introduce the electron operators at a given point in real space,
for the right- and left-moving electrons, in the form
\begin{align}
\psi_{\sigma +}(x) = R_\sigma(x) e^{i k_F x}, \quad \psi_{\sigma -}(x) = L_\sigma(x) e^{-i k_F x},
\end{align}
where $R_\sigma(x)$ and $L_\sigma(x)$ vary slowly on the scale of $k_F^{-1}$.
Specifically, $\mathcal{H}_0$ and $\mathcal{H}_f$ are written similar to the conventional Luttinger model as
\begin{align}
\mathcal{H}_0 = -i v  \left( R_{\sigma}^{\dagger} \partial_x R_{\sigma}^{}- L_{\sigma}^{\dagger} \partial_x L_{\sigma}^{}  \right)
\end{align}
and
\begin{align}
\begin{split}
\mathcal{H}_f =& \enspace  \sum_{\sigma\sigma'}\left(g_{2\parallel} \delta_{\sigma\sigma'}
+ g_{2\perp} \delta_{\sigma,-\sigma'}\right) \rho_{R\sigma} \rho_{L\sigma'} \\
& +  \frac{1}{2} \sum_{\sigma\sigma'\eta}\left(g_{4\parallel} \delta_{\sigma\sigma'}
+ g_{4\perp} \delta_{\sigma,-\sigma'}\right) \rho_{\eta\sigma} \rho_{\eta\sigma'}~,
\end{split}
\end{align}
where the chiral densities in $\mathcal{H}_f$ are given by
$\rho_{R\sigma} = R_{\sigma}^{\dagger} R_{\sigma}^{\phantom{\dagger}}$ and $\rho_{L\sigma}
= L_{\sigma}^{\dagger} L_{\sigma}^{\phantom{\dagger}}$, and the coupling constants read
$g_{4\parallel} = g_{2\parallel} = V_{11}(0)$ and $g_{4\perp} = g_{2\perp} =V_{12}(0)$. The backscattering part
\begin{align}
\begin{split}
\mathcal{H}_b =&  \sum_{\sigma\sigma'} \left( g_{1\parallel} \delta_{\sigma\sigma'}
+ g_{1\perp} \delta_{\sigma,-\sigma'} \right) h^{\dagger}_{\sigma} h_{\sigma'}
\label{Hbhh}
\end{split}
\end{align}
is represented in terms of $R_\sigma$ and $L_\sigma$ differently, compared to the conventional Luttinger model,
with $h_\sigma$ coming from the spatial gradient expansion:
\begin{align}
h^{\phantom{\dagger}}_{\sigma} =  \left[\,(\partial_x R^{\dagger}_{\sigma}) L_{\sigma}^{\phantom{\dagger}}
- R^{\dagger}_{\sigma} (\partial_x L_{\sigma}^{\phantom{\dagger}})\,\right]a~,
\end{align}
where $a$ is the ultraviolet cutoff in real space. As discussed above, the local representation of the backscattering term
is valid on spatial scales larger than $d$, hence $a\sim d$.
The coupling constants for backscattering are given by
\begin{equation}
g_{1\parallel}=\frac{4k_F^2}{k_0^4a^2}V_{11}(2k_F), \quad g_{1\perp}=\frac{4k_F^2}{k_0^4a^2}V_{12}(2k_F).
\label{g1}
\end{equation}
Note that the ultraviolet scale $a$ cancels out in Eq.~(\ref{Hbhh}).

The forward scattering term $\mathcal{H}_f$ can be treated exactly by bosonization,
with the fermionic fields represented in terms of the bosonic field $\varphi_{\sigma}(x)$ and its canonical
conjugate $\theta_{\sigma}(x)$ as
\begin{align}
R_{\sigma}(x) =& \,\frac{1}{\sqrt{2 \pi a}}\, e^{i \sqrt{\pi} [ \varphi_{\sigma}(x) - \theta_{\sigma}(x)] } \, ,\\
L_{\sigma}(x) =& \,\frac{1}{\sqrt{2 \pi a}}\, e^{-i \sqrt{ \pi}[ \varphi_{\sigma}(x) + \theta_{\sigma}(x)] }\, .
\end{align}
Changing from the ``wire basis" ($\sigma= 1,2$) to the basis of symmetric (+) and antisymmetric ($-$) fields
\begin{equation}
\varphi_\pm = (\varphi_1\pm\varphi_2)/\sqrt{2},
\quad \theta_\pm=(\theta_1\pm\theta_2)/\sqrt{2},
\end{equation}
the bosonized Hamiltonian density reads
\begin{align}
\begin{split}
\mathcal{H} =& \sum_{\lambda=\pm}
\frac{v_{\lambda}}{2}\left[\,K_{\lambda}(\partial_x\theta_\lambda)^2+K_{\lambda}^{-1}(\partial_x\varphi_{\lambda})^2\,\right]\\
+& \;\frac{g_{1\perp}}{\pi}\left[\,(\partial_x\theta_+)^2-(\partial_x\theta_-)^2\,\right]\cos\left(2\sqrt{2\pi}\varphi_{-}\right)~,
\label{bosonic model}
\end{split}
\end{align}
where
\begin{align}
K_{\pm}= &\,\,\sqrt{\frac{1-U_{\pm} }{1+ U_{\pm}}}~,
\label{K-}\\
v_{\pm} =&\,\,u_{\pm} \sqrt{1- U^2_{\pm}   } \vphantom{\frac{1}{1}}
\label{plasmonvelocities}
\end{align}
with
\begin{align}
U_\pm=&\,\,\frac{1}{2 \pi u_\pm}\left(g_{2\parallel}\pm g_{2\perp}\right)~,
\label{U-}\\
u_{\pm} =&\,\,v+ \frac{1}{2 \pi }\left(g_{4\parallel}\pm g_{4\perp}\right)~.
\end{align}
Note that for $g_{2\parallel}\pm g_{2\perp} =g_{4\parallel}\pm g_{4\perp}$, the
relation $v_{\pm}K_{\pm}=v$ holds.

In the bosonized Hamiltonian (\ref{bosonic model}), we have disregarded the terms arising due to the
intraedge backscattering ($g_{1\parallel}$-terms), since they contain the fourth power of gradients and hence
are highly irrelevant in the infrared. This should be contrasted with the conventional Luttinger liquid,
where such terms can be fully incorporated by shifting $g_{2\parallel}$.

Because of the $g_{1\perp}$ scattering processes, the coupling constants for
the double-edge system described by the Hamiltonian (\ref{bosonic model}) are subject to renormalization.
Under a renormalization-group (RG) transformation that keeps the quadratic term in Eq.~(\ref{bosonic model})
invariant, the scaling dimension for the backscattering operator is obtained as $2 (K_--1) +2$, with $2 (K_--1)$
describing the renormalization of the cosine term and the second term coming from the spatial gradients.
That is, backscattering is irrelevant in the RG sense, with the coupling constant
\begin{equation}
\alpha_b=\frac{g_{1\perp}}{2\pi v_-}
\end{equation}
scaling with $T$ as
\begin{align}
\alpha_b(T)=\alpha_{b0}\left(\frac{Td}{v}\right)^{2K_-} ~,
\end{align}
where $\alpha_{b0}$ in the bare coupling.

It is worth noting that the renormalization of $g_{1\perp}$ here is governed by the forward-scattering amplitudes
encoded in $K_-$,
in contrast to the conventional spinful Luttinger liquid, where the backscattering is renormalized by itself
(more precisely, the weak-coupling beta function for $g_{1\perp}$ is proportional to the product
$g_{1\perp}g_{1\parallel}$) \cite{Giamarchi_book}.
The difference stems from the inequality $g_{2\perp}\neq g_{2\parallel}$ that gives $K_-<1$ for spatially distant edges
and from the strong irrelevance of the $g_{1\parallel}$ interaction.

We will return to the scaling behavior of ``first-order backscattering" in Sec.~\ref{Sec:Luttinger liquid renormalization},
when calculating $\rho_D$. For now, we proceed with the RG treatment of Eq.~(\ref{bosonic model}).
The discussion above brought up an important point that the theory with $\mathcal{H}$ from Eq.~(\ref{bosonic model})
is weakly coupled, {\it provided} no additional couplings that become relevant are generated by the RG transformation.
In fact, as we discuss in Sec.~\ref{Sec:Renormalization group analysis}, second-order backscattering processes
do become relevant for sufficiently strong forward-scattering interactions.

\subsection{Higher-order backscattering}
\label{Sec:Renormalization group analysis}

As already mentioned in Sec.~\ref{s4}, the backscattering operator in Eq.~(\ref{bosonic model}),
which is itself irrelevant, can generate relevant operators under the RG transformation.
These describe higher-order backscattering processes. Among the additional backscattering terms in the rescaled Hamiltonian,
the relevancy is the highest for the term proportional to the next-order, compared to Eq.~(\ref{bosonic model}),
harmonic of the field $\varphi_-$, i.e., to $\cos (4\sqrt{2\pi}\varphi_-)$.
Importantly, the emergent additional backscattering interaction is not suppressed by spatial gradient terms in
the prefactor of the cosine, in contrast to Eq.~(\ref{bosonic model}).
Specifically, as shown in Appendix~\ref{App:Derivation of higher order backscattering processes},
the perturbative (in $\alpha_b\ll 1$) RG yields the $\cos (4\sqrt{2\pi}\varphi_-)$ term at the second order in $g_{1\perp}$.
The resulting effective action reads $S=S_0+S_1+S_2$, where
\begin{align}
\begin{split}
S_0=& \sum_{\lambda=\pm}
\int \! \mathrm{d} x \mathrm{d} \tau \,
  \left\{-i\partial_{x} \theta_\lambda \partial_\tau \varphi_\lambda \right. \\
+&
\left.\frac{v_\lambda}{2} \left[\, K_\lambda(\partial_{x} \theta_\lambda)^2 + \frac{1}{K_\lambda}(\partial_x \varphi_{\lambda})^2\,\right]\right\}~, \\
S_1=& 2 v_-\alpha_b\!
 \int \! \mathrm{d} x \mathrm{d} \tau\left[(\partial_x\theta_+)^2-(\partial_x\theta_-)^2\right] \cos\left(2\sqrt{2\pi}\varphi_{-}\right),\\
             S_2=&  v_- \beta_b\int \! \frac{\mathrm{d} x \mathrm{d} \tau}{\pi a^2} \,  \cos\left(4\sqrt{2 \pi} \varphi_-\right)~.
\label{RGaction}
\end{split}
\end{align}
Here we discarded the highly irrelevant terms stemming from $g_{1\parallel}$ that modify $S_0$ by introducing terms with higher gradients
(in this regard, their effect is similar to the effect of a finite curvature of the electronic dispersion relation).
Note that the term $S_1$ couples the antisymmetric ($-$) sector with the symmetric ($+$) one, but, as discussed above,
this term is irrelevant in the RG sense (at least, in the weak-coupling regime).
The structure of the term $S_2$ suggests its interpretation as describing the processes of correlated four-fermion backscattering.
A similar term with doubled harmonics is generated under the RG in disordered helical edges, see, e.g., Ref.~\cite{Kainaris_2014},
where it described a two-particle backscattering off the random potential.

Neglecting $S_1$, the action (\ref{RGaction}) becomes identical to that for two coupled spinless TLLs,
characterized by the Luttinger constant $K_-^{\text{TLL}}$ for the antisymmetric field $\varphi_-^{\text{TLL}}$,
if one changes $K_-\to K_-^{\text{TLL}}/4$ and rescales $\varphi_-\to\varphi_-^{\text{TLL}}/4$.
One important consequence of this mapping is that the system of two strongly correlated helical liquids with $K_-\simeq 1/4$
behaves similarly to weakly interacting TLLs. In particular, there is a Berezinskii-Kosterlitz-Thouless (BKT) transition
in the limit of $g\to 0$ at $K_-=1/4$ \cite{Giamarchi_book}.
Specifically, the RG equations for the coupling constants $K_-$ and $\beta_b=g/2\pi v_-$,
which characterize the action (\ref{RGaction}), read
\begin{align}
& \frac{dK_-}{d\ell}=-8\beta_b^2K_-^2~, \qquad \frac{d \beta_b}{d \ell} = 2(1-4K_-)\beta_b~,
\label{BKTRG}
\end{align}
where $\ell = \ln (\Lambda_d / \Lambda)$, the ultraviolet cutoff $\Lambda_d$ in energy space is of the order of $v/d$, and $\Lambda$
is the running cutoff.

\begin{figure}
  \centering
  \includegraphics[width=.3\textwidth]{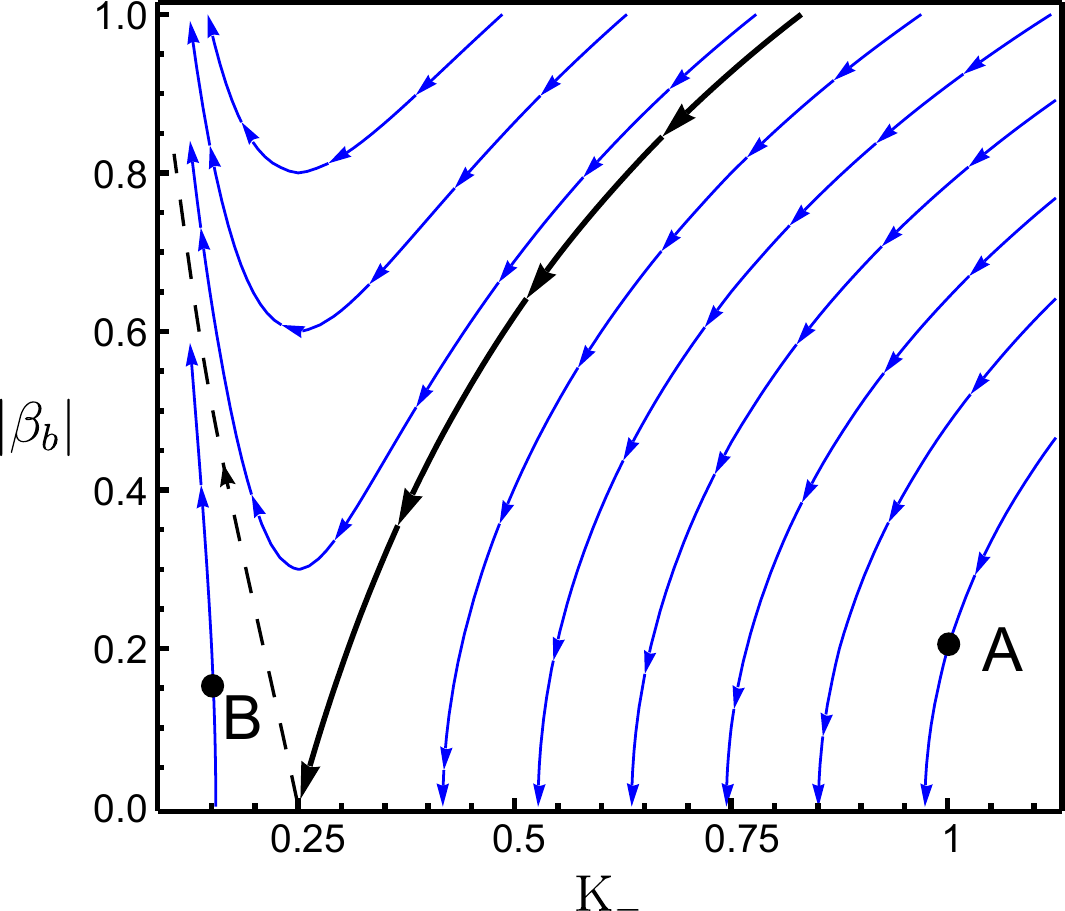}
  \caption{Renormalization-group flow of the coupling constant for interwire second-order backscattering $\beta_b$ vs
  the Luttinger constant for the relative charge mode $K_-$.
  The separatrix between the strong- (point B at $K_-<1/4$) and weak-coupling (point A at $K_-\simeq 1$) phases (thick black line)
  terminates at a strongly-interacting point with $K_-=1/4$. To the left of the dashed curve the magnitude of $\beta_b$ grows monotonously.
  }
  \label{Fig:RGflow}
\end{figure}

The bare value of $\beta_b$ in Eq.~(\ref{BKTRG}) is always smaller than the bare value of $\alpha_b$.
Indeed, on the ultraviolet scales $\Lambda\sim\Lambda_d$, the coupling constant $\beta_{b0}$ is quadratic in
$g_{1\perp}$ and proportional to the function $F(K_-, K_+)$ from Eq.~(\ref{FKK}).
Importantly, the function $F(K_-, K_+)$ is nonzero for $K_\pm>0$, so that the second-order backscattering is always generated.
According to the weak-coupling RG equations (\ref{BKTRG}), the sign of $\beta_{b}$ is not changed in the course
of the renormalization, while the renormalization of $K_-$ is insensitive to the sign of $\beta_b$ by Eq.~(\ref{BKTRG}).
Furthermore, for $|\beta_b|\ll 1$, inelastic processes mediated by second-order backscattering
(see Sec.~\ref{Sec:Luttinger liquid renormalization} below) are also insensitive to the sign of $\beta_b$.
Therefore, in what follows, when discussing the weak-coupling regime, we will use the notation $\beta_b$ for $|\beta_b|$.

The integral curves
\begin{align}
\beta_b(K_-)=\left[\,\beta^2_{b0}+2\left(\frac{1}{4K_-}-\frac{1}{4K_{0}}-\ln\frac{K_0}{K_-}\right)\,\right]^{1/2}
\end{align}
of the RG flow for different initial conditions $\beta_{b0}$ and $K_0$ are shown in Fig.~\ref{Fig:RGflow}.
The separatrix
\begin{align}
\beta_b^{(s)}(K_-)=\left[\,2\left(\frac{1}{4K_-}-1-\ln\frac{1}{4K_-}\right)\,\right]^{1/2}
\label{separatrix}
\end{align}
divides the phase space into the basin of attraction for the line of weak-coupling fixed points with $\beta_b=0$ and $K_->1/4$
(by way of illustration, point A in Fig.~\ref{Fig:RGflow})
and the region in which the flow is to strong coupling (growing $\beta_b$ with $K_-<1/4$, point B in Fig.~\ref{Fig:RGflow})).

Note that the behavior of $\beta_b$ as a function of $T$ reduces, for $\beta_{b0}\ll 1$, to a simple power law (one can neglect the renormalization of $K_-$):
\begin{align}
\beta_b(T)=\beta_{b0}\left(\frac{Td}{v}\right)^{8K_0-2}~.
\end{align}
For $K_0<1/4$, second-order backscattering becomes stronger as $T$ is decreased.
The characteristic temperature $T^\ast$ at which Eq.~(\ref{BKTRG}) gives
$\beta_b (T^\ast)\sim 1$ corresponds to the onset of the strong-coupling regime.
For the interwire potential specified below Eq.~(\ref{elementsB}), the result for this temperature scale is given by
\begin{align}
T^{\ast}\sim \frac{v}{d}\,\beta_{b0}^{1/(2-8K_-)}\propto \exp \left(-\dfrac{k_F d}{1-4 K_-}\right)~.
\label{3}
\end{align}
with $K_-=K_0$.

In the strong-coupling limit, the term $S_2$ in the action (second-order backscattering) tends to lock the phase $\varphi_-$ at the minima
of the cosine potential in Eq.~(\ref{RGaction}), which implies the formation of a charge-density wave in
the double-wire system.
This is similar to the strong-coupling regime for two conventional spinless Luttinger liquids with repulsive
interactions \cite{Nazarov_Averin_1998,Klesse_Stern_2000}. At the level of the action $S_0+S_2$, the difference is
that there is a threshold value for the strength of repulsive interactions below which the strong-coupling regime
cannot be reached in helical liquids,
whereas arbitrarily small repulsion between electrons drives the system into the strong-coupling regime
in conventional Luttinger liquids.
We will discuss Coulomb drag between helical liquids for the case of strong coupling
in Sec.~\ref{Sec:Strong coupling drag}.

\subsection{Luttinger-liquid renormalization of the drag resistivity}
\label{Sec:Luttinger liquid renormalization}

In this Section, we calculate the drag resistivity by
incorporating the power-law renormalization which is characteristic of the Luttinger-liquid physics.
The effect of forward scattering ($K_-<1$) on Coulomb drag mediated by the first-order backscattering
(described by the term $S_1$ in the action) can be taken into account by using a renormalized
interaction coupling constant $\alpha_b(T)$ in the results obtained above by means of
solving the kinetic equation (Sec.~\ref{Sec:Coulomb drag in helical edges}).
This amounts to the replacement
\begin{equation}
V_{12}\to V_{12}\left(\frac{Td}{v}\right)^{2(K_--1)}
\end{equation}
in the results of Sec.~\ref{Sec: Low frequency drag}.
In particular, for $T \to 0$ this replacement yields
\begin{align}
   \rho_D \sim \frac{1}{e^2}\alpha_{b}^2(T)\, T \sim
   \frac{1}{e^2} \alpha_{b0}^2 \left(\frac{Td}{v} \right)^{4K_-}\!\!  T \propto T^{4K_-+1}.
   \label{scalingrho}
\end{align}
The dependence of the prefactor of the power law on $K_-$ in this
expression is obtained in Appendix~\ref{App:Renormalization}.

As pointed out in Sec.~\ref{Sec:Renormalization group analysis},
the backscattering operator in Eq.~(\ref{bosonic model}),
which itself is irrelevant, can generate relevant operators under the RG flow.
These describe higher-order backscattering
processes that contribute to the drag resistivity at order $\alpha_{b0}^4$.
As we will see below, one cannot neglect these
contributions, even though they are of higher order in the bare
interedge interaction strength. This is because they may scale
with a lower power of $T$ than the first-order backscattering contributions
and hence may overcome the lowest-order at sufficiently
low $T$. Moreover, for the case of sufficiently strong interaction
(low values of $K_-$), these contributions lead to the increase
of $\rho_D$ with lowering $T$.

The effect of second-order backscattering is not captured by the kinetic-equation
approach developed above, which includes
only pair collisions and neglects interedge correlations. To obtain the drag
conductivity due to the second-order backscattering
processes, we calculate the drag conductivity using the Kubo formula,
\begin{align}
    \sigma_D(\omega) = - \frac{i}{\omega} \lim_{i \Omega_n \to \omega +i 0}
    \lim_{q \to 0} \braket{j_1(q,\Omega_n) j_2(-q,-\Omega_n)} \, ,
 \end{align}
where $j_\sigma(x,\tau) = e K_{\sigma} v_{\sigma} \partial_x \theta(x,\tau) / \sqrt{\pi}$
is the current in edge $\sigma =1,2$.
Here, the correlation function is calculated with respect to the action
$S = S_0 + S_2$ in Eq.~(\ref{RGaction}), yielding the high-frequency drag conductivity
\begin{multline}
   \text{Re}\,\sigma_D(\omega) = \frac{e^2 v_-}{\omega^2}
   \beta_{b0}^2 \left(\frac{\pi Td}{v_-} \right)^{16 K_- -3} \frac{v_-}{d} \Upsilon(K_-) \, , \\
   \Upsilon(K_-) =  \frac{8}{\pi^2}\cos^2(4 \pi K_-)
   \Gamma^2(\textstyle{\frac{1}{2}} - 4K_-) \Gamma^2( 4K_-) \, , \label{Luttingerdrag2}
\end{multline}
where $\Gamma(x)$ is the Euler gamma function.

The total high-frequency drag conductivity is a sum of the contribution due to
first-order backscattering and the contribution of Eq.~(\ref{Luttingerdrag2}).
Following the reasoning presented in Sec.~\ref{Sec: Low frequency drag}, we
expect that the dc drag resistivity is determined by the high-frequency drag rate
extracted from the ac conductivity. A rigorous analysis of the renormalized dc
drag resistivity can be performed in a two-step way. First, one renormalizes
the bosonized theory down to the energy scale given by $T$.
Second, one refermionizes the theory and solves the kinetic equation for the new
fermionic excitations. We relegate this program to future work.
Comparing Eq.~(\ref{Luttingerdrag2}) with Eq.~(\ref{Luttingerdrag1}),
we see that the second-order contribution scales with a lower power of $T$ when
the interedge correlations are sufficiently strong, $K_- <1/3$.
Moreover, as we have already shown, second-order backscattering becomes
relevant for $K_- <1/4$. Then, the weak-coupling analysis performed
above is only valid for sufficiently high temperatures.

Summarizing, the low-$T$ drag resistivity is dominated by the first-order
backscattering processes, Eq.~(\ref{Luttingerdrag1}), as long as $K_->1/3$.
For stronger repulsive intraedge interaction, $1/4<K_-<1/3$, the low-$T$
drag resistivity is governed by second-order backscattering, Eq.~(\ref{Luttingerdrag2}).
At $K_-<1/4$, these scattering processes become relevant in the RG sense
and lead to the increasing drag resistivity as $T$ is lowered.

\subsection{Coulomb drag in the strong-coupling limit}
\label{Sec:Strong coupling drag}
In this section, we discuss the drag resistivity at the strong-coupling fixed point of the RG flow
derived in Sec.~\ref{Sec:Renormalization group analysis}.
We remind the reader that there is a mapping of the bosonized theory described by the action $S_0+S_2$
(neglecting the irrelevant term $S_1$) in Eq.~(\ref{RGaction})
to the theory of coupled TLLs with the Luttinger constant $K=4K_-$.
Thus, the problem of Coulomb drag in helical edges in the strong-coupling regime is expected
to bear similarity to the drag between spinless TLLs discussed in Refs.~\cite{Nazarov_Averin_1998,Klesse_Stern_2000}
and to the problem of pinned charge density waves~\cite{Rice_1976,Maki_1977}.
To keep our analysis self-contained, we will reproduce here the main results of these works,
using the notation of Sec.~\ref{Sec:Renormalization group analysis} and only keeping the terms $S_0$ and $S_2$
in the bosonized action.

For definiteness, we assume that $\beta_b$ is positive
(for negative $\beta_b$ the consideration is qualitatively similar).
In the strong coupling limit, with $\beta_b \gg 1 $ and $K_- \ll 1/4$, the action $S_0+S_2$
is minimized by the uniform mean-field configurations
$\sqrt{32 \pi} \varphi_-(x) = \phi_m \equiv  (2 m +1) \pi$.
Recall that $\varphi_-$ describes the displacement of electrons in wire 2 with respect to electrons in wire 1, so that
the mean-field solution gives two interlocked charge density waves. At finite $T$,
there exist massive harmonic fluctuations around this mean-field
result. However, these excitations do not carry the antisymmetric current.

Electron transport from one end of the active wire to the other only occurs if the mean-field value of
the field changes from $\phi_m$ to $\phi_{m\pm1}$. Depending on temperature, transitions between the degenerate
ground states occur due to either quantum tunneling or thermal activation.
The excitations that carry the current are either \mbox{(anti-)solitons} that move along the wire
or soliton-antisoliton pairs that are formed inside the wire and dissociate by the applied electric field.
The energy $E_s$ and width $W_s$ of a classical soliton are \cite{Maki_1977}
\begin{align}
   E_s = \sqrt{\frac{2 \beta_b}{\pi^2 K_-}}  \frac{v_-}{a}\, , \qquad W_s = \frac{1}{4\sqrt{ K_- \beta_b }} a \, .
\end{align}
When $W_s$ is much smaller than the system length,
the response in the antisymmetric sector is determined by the thermal excitation
of soliton-antisoliton pairs and reads as \cite{Rice_1976}
\begin{align}
   \rho_{\text{therm}} = \frac{h}{32 \pi e^2 \ell_s} \sqrt{\frac{E_s T a^2}{2 \beta K_- v_-}} e^{E_s/T} \, ,
   \label{thermalresistivity}
\end{align}
where $\ell_s$ is the soliton mean free path.

In conventional TLLs, repulsive backscattering interactions between two (spinless) wires always become stronger as $T$ is
decreased---for arbitrary bare strength of the interactions. That is, there exists a characteristic temperature $T^\ast$ at
which the coupling constant for electron-electron backscattering
becomes of the order of unity. Below $T^\ast$, a zigzag-ordered charge density wave is formed, as discussed above.
As a consequence, one of the primary properties of Coulomb drag between TLLs with repulsive interwire
interactions is that $\rho_D$ shows activation behavior, Eq.~(\ref{thermalresistivity}), for $T\ll T^\ast$:
\begin{align}
\ln\dfrac{\rho_D(T)}{\rho_D(\Delta)}\simeq\dfrac{\Delta}{T}
\label{2}
\end{align}
with the activation gap $\Delta \sim T^{\ast}$ \cite{Nazarov_Averin_1998}.

Based on the above mapping, one concludes that for $K_-<1/4$ the drag
resistivity has a local minimum at a characteristic temperature
$T^\ast$, Eq.~(\ref{3}), at which the coupling constant $\beta_b$
for the second-order backscattering becomes of the order of unity.
Below this temperature, $\rho_D(T)$ starts growing exponentially
due to thermally activated transitions between neighboring
ground states. On the other hand, if $K_->1/4$,
this local minimum does not occur and the drag resistivity vanishes as a power law as $T \to 0$.

We emphasize that the above conclusion is based on retaining only the terms $S_0$ and $S_2$
in the bosonized action (\ref{RGaction}).
If the mapping onto the conventional theory of 1D Coulomb drag worked for the helical edges,
for $K_-<1/4$ one would obtain $\rho_D=-\rho_{12}\to \infty$ at $T\to 0$.
However, from the general structure of the resistivity tensor, it follows that the diagonal
(intraedge) resistivity should diverge simultaneously: $\rho_{11}\to \infty$. Indeed,
for clean (no disorder) systems we have 
$\rho_{11}=-\rho_{12}$ \footnote{Note that  the contribution of the
$g_{5}$-processes  to $\rho_{11}$ vanishes at $T\to 0$ \cite{Kainaris_2014}.}.
Thus, the divergence of the drag resistivity would mean that the interedge coupling
destroys the topological protection of the (otherwise) conducting helical edge states.
Specifically, on both sides of the quantum spin-Hall transition driven by the closing
and reopening of the gap in the 2D bulk of the system (gap inversion), we would then
have nonconducting edge states. However, at zero gap, the 2D bulk state is still conducting;
therefore, the delocalized bulk state is expected to transform into the
conducting edge state at one side of the QSH transition.

We speculate that, within the framework of an effective edge theory, this ``topological protection'' might be
related to the difference between the HLL and normal TLL: the former contains the additional term $S_1$ in
the action (``$\alpha$-term"). While in the weak-coupling regime this term is highly irrelevant, when
the ``$\beta$-term" $S_2$ enters the strong-coupling limit, the $\alpha$-term might again become important,
destroying the charge density wave. In this scenario, the topological
protection is maintained due to the competition of the $\alpha$ and $\beta$ terms in the action, leading to
nonperturbative effects in the strong-coupling regime. A somewhat similar situation was encountered in
Ref.~\cite{OGM10} devoted to the 2D surface states of a 3D topological insulator.
There, the perturbative (weak-coupling) RG suggested a localization of the surface states,
due to the Altshuler-Aronov-type corrections, but the nonperturbative effect of the
topological protection resulted in the emergence of a critical state in the strong-coupling regime
\footnote{In this paper, we restrict ourselves to the analysis of the RG equations derived at the
lowest order in the couplings $\alpha_b$ and $\beta_b$. The next-order terms in the beta functions
might give rise to a precursor of the topological protection already in the weak-coupling regime (for $K_-\ll 1$).}.

To conclude this section, the behavior of $\rho_D$ for sufficiently strong intraedge interaction, namely
$K_-<1/4$, is expected to be governed by the interplay of the tendency to the formation of a charge density wave
because of second-order backscattering (the term $S_2$ in the action) on the one hand and the topological protection (encoded in
the $\alpha$-term $S_1$) on the other. This interplay might lead to a nonmonotonic $T$ dependence of $\rho_D$
at low $T$, with a local minimum around $T^\ast$ and a local maximum at yet lower $T$.
The behavior of $\rho_D$ at finite $T$ would then demonstrate an ``apparent metal-insulator transition''
with
decreasing $K_-$. Based on the weak-coupling RG formalism, we cannot make definite conclusions about the nature
of zero-$T$ Coulomb drag for $K_-<1/4$. We relegate the corresponding analysis to future work.

\section{Summary}
\label{Sec: Summary}

We have presented a theory of Coulomb drag between clean (no disorder) helical Luttinger liquids based on
the kinetic equation approach supplemented with bosonization to take into account Luttinger liquid renormalization.
We have assumed that the spin-rotational invariance of the helical liquid is broken by Rashba spin-orbit coupling in the
bulk of the topological insulator, which allows for interedge backscattering events without breaking time-reversal
symmetry. We have obtained a richer phase diagram for Coulomb drag in helical liquids compared to conventional 1D wires with
repulsive interactions.

A peculiar feature of Coulomb drag between helical
liquids---related to the existence of the Dirac point---is exposed in the high-$T$ case.
We have shown that Coulomb drag between helical liquids is mediated not only by backscattering but also by Umklapp processes.
In the helical liquid, Umklapp scattering is special in that the energy and momentum conservation makes it necessary
for one of the involved states (either initial or final) of Umklapp-scattered particles to be right at the Dirac point
(see Fig.~\ref{Fig:scatteringprocesses}).
The Umklapp processes reveal themselves in Coulomb drag between helical liquids in a subtle manner.
Their direct contribution to the drag resistivity $\rho_D$, actually, vanishes exactly; nonetheless, they impact Coulomb drag profoundly
by providing for coupling to plasmon modes. In turn, Coulomb drag is dominated by the excitation of plasmons,
triggered by Umklapp scattering. This results in a nonmonotonic $T$ dependence of $\rho_D$,
characterized by several crossovers.
In particular, in the limit of high $T$, the drag resistivity falls off in a universal manner as
\begin{equation}
 \rho_D\propto\,\alpha_{b0}^2\, T^{-4},
\end{equation}
where $\alpha_{b0}$ describes the strength of interedge backscattering.

In helical liquids, backscattering is much weakened by spin-momentum locking, and one important question pertinent
to the behavior of $\rho_D$ in the low-$T$ limit, where
Luttinger-liquid effects become important, is about the outcome of a competition between strong correlations
and the spin-momentum locking. The impact of the Luttinger renormalization on the drag resistivity is twofold.
First, it renormalizes the power-law exponents in the temperature dependence of $\rho_D$, depending on
the strength of interactions characterized by the Luttinger parameter $K_-$ of the relative charge mode.
Second, for $K_-<1/4$, higher-order electron-electron backscattering processes become strong below a characteristic
temperature scale $T^{\ast}$ and tend to form a charge density wave.
We have shown that, if repulsive interactions are not too strong, namely $K_->1/4$, the spin-locking wins and $\rho_D$
vanishes at $T\to 0$ as a power law of $T$:
\begin{align}
\rho_D\propto\,\left\{
                           \begin{array}{ll}
                             \alpha_{b0}^2\left(\frac{T}{T_d}\right)^{4K_-+1}~, & K_->1/3~, \\
                             \alpha_{b0}^4\left(\frac{T}{T_d}\right)^{16K_--3}, & 1/3>K_->1/4~.
                           \end{array}
                         \right.
\label{4}
\end{align}
For $K_-<1/4$, the system enters the strong-coupling regime at $T\alt T^\ast$.
We expect a nonmonotonic behavior of $\rho_D$ as $T$ is lowered further,
governed by the interplay between the formation of the charge density wave on the one hand and the topological protection on the other, with a local minimum in the $T$ dependence of $\rho_D$ at $T\sim T^\ast$
and a local maximum at yet lower $T$.

\begin{figure}
  \centering
  \includegraphics[width=.4\textwidth]{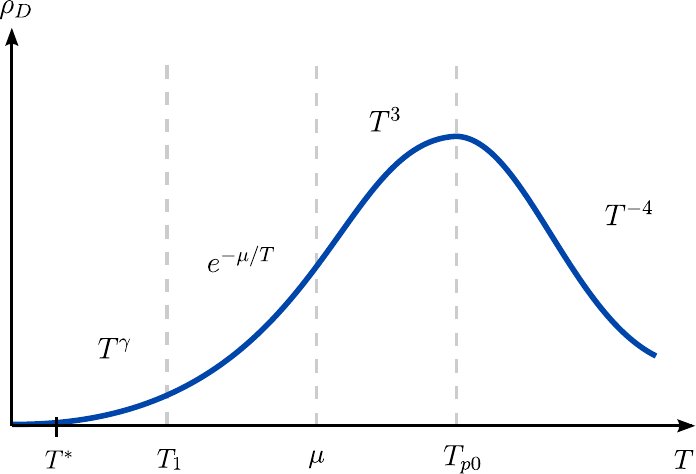}
  \caption{
  Sketch of the $T$ dependence of the drag resistivity $\rho_D$ for two helical liquids for $v/d\ll |\mu|\ll  T_{p0}$,
  where $T_{p0}$ [Eq.~(\ref{e16})] is the temperature above which the plasmon damping leads to a strong suppression
  of the drag rate. For $T\gg T_1$ [Eq.~(\ref{51})], electron-hole excitations in two edges couple to
  each other through plasmon modes (``plasmon-mediated Coulomb drag"). For $T \ll T_1$, Coulomb drag is
  determined by direct backscattering of electrons close to the Fermi surface. The exponent $\gamma$ in the
  power-law $T$ dependence of $\rho_D$ at $T\to 0$ is given by $\gamma=4K_-+1$ for $K_->1/3$ and $\gamma=16K_--3$ for $1/4<K_-<1/3$.
  For $K_-<1/4$, as $T$ decreases, the system enters the strong-coupling regime at $T\sim T^*$
  (Sec.~\ref{Sec:Renormalization group analysis}).
    }
  \label{Fig:dragplot1}
\end{figure}

Summarizing, the overall picture of the dependence of $\rho_D$ on $T$,
as follows from the results of Secs.~\ref{sIIId}--\ref{Sec:Strong coupling drag},
is illustrated in Fig.~\ref{Fig:dragplot1}.
Viewed from a general perspective, Fig.~\ref{Fig:dragplot1} demonstrates that Coulomb drag between helical liquids
is, as already emphasized above, peculiar in two important aspects.
One of the peculiarities, apparent in Fig.~\ref{Fig:dragplot1},
is that $\rho_D$ vanishes with decreasing $T$ as a power law if intraedge interactions are not too strong.
The other, highly unusual, property of helical liquids that we explored in this paper is the peculiar
Umklapp-triggered plasmon-mediated mechanism of
Coulomb drag that governs the behavior of $\rho_D$ for higher temperatures.
This behavior clearly distinguishes Coulomb drag in helical
liquids from drag in conventional 1D quantum liquids and could be used to identify helical liquids in Coulomb-drag experiments
\footnote{Experimentally, the study of Coulomb drag between helical liquids can be performed
either with the vertical setup consisting of two parallel quantum wells, as shown schematically in Fig.~\ref{Fig:dragsetup},
or with the horizontal setup (C. Br{\"u}ne and H. Buhmann, private communication),
where the two quantum wells are located in the same plane.}.

\section{Acknowledgements}
\label{Sec: Acknowledgements}

This work was supported by the DFG SPP 1666 ``Topological insulators'' and
by the EU Network FP7-PEOPLE-2013-IRSES under Grant No. 612624 ``InterNoM."
NK thanks the Carl-Zeiss-Stiftung for financial support. The work at University of Wisconsin-Madison was financially supported in part by
NSF Grants No. DMR-1606517, No. ECCS-1560732, and by the Wisconsin Alumni Research Foundation.
We acknowledge discussions with C.~Br\"une, H.~Buhmann, S.T.~Carr, Y.-Z.~Chou, L.~Du, M.~Foster, A.D.~Mirlin, and B.~Trauzettel.

\appendix

\section{Polarization operator and RPA interaction for the helical edge states}
\label{a1}

In this Appendix, we derive the polarization operator and the dynamically screened RPA interaction
for a homogeneous HLL. The density of helical fermions in edge $\sigma$ is written as
\begin{align}
   \rho_{\sigma}(q) =  \sum_{\eta_1\eta_2 }\int_k \,  {\psi}^\dagger_{k+q,\eta_1\sigma} \psi_{k\eta_2\sigma} b_{\eta_1\eta_2}(k+q,k) \, ,
\end{align}
with the matrix elements $b_{\eta_1\eta_2}(k_1,k_2)$ defined in Eq.~(\ref{elementsB}).
The (bare) polarization operator in the Matsubara representation,
\begin{align}
   \Pi(q, i \Omega_m) = - \braket{\rho(q,i \Omega_m) \rho(-q,-i \Omega_m) }
\end{align}
(given that we have identical edges and no tunneling between them, the index $\sigma$ is dropped here and below),
averaged over the noninteracting ground state, is a sum $\Pi=\sum_{\eta_1\eta_2}\Pi_{\eta_1\eta_2}$ of the chiral components
\begin{multline}
   \Pi_{\eta\eta}(q, i \Omega_m)\\ = -  T \sum_n \int_k\, G_{0\eta}(k+q,i\omega_n + i \Omega_m) G_{0\eta}(k, i \omega_n)
\end{multline}
and the backscattering components
\begin{multline}
   \Pi_{\eta, -\eta}(q, i \Omega_m) = -  T \sum_n \int_k\, G_{0,-{\eta}}(k+q,i\omega_n + i \Omega_m)\\
   \times  G_{0\eta}(k, i \omega_n)  b_{\eta, -\eta}(k,k+q) b_{-\eta, \eta }(k+q,k) \, ,
\end{multline}
where the bare fermion propagator reads
\begin{align}
   G_{0\eta}(k) = (-i\omega_n + v \eta k - \mu)^{-1} \, .
\end{align}
We have, then,
\begin{align}
    \Pi_{\eta\eta}(q, i \Omega_m) = \frac{1}{2 \pi v } \frac{v\eta q}{v\eta q - i \Omega_m}
\end{align}
and
\begin{multline}
   \Pi_{\eta,-\eta}(q, i \Omega_m) =
   - \frac{1}{k_0^4} \int_k \, \frac{\left[ k^2-(k+q)^2\right]^2}{i \Omega_m + 2 v\eta k +  v \eta q}
   \\    \times
   \left[\, n_F(v \eta k) - n_F(-v \eta k - v \eta q) \,\right] \, .
\end{multline}
\begin{widetext}
After the analytical continuation to real frequencies $i \Omega_m \to \Omega + i 0$, the retarded backscattering
polarization operator $\Pi_{\eta,-\eta}(q,\Omega)$ is given by
\begin{align}
\begin{split}
 & \text{Re}\,\Pi_{\eta,-\eta}(q,\Omega) = - \frac{1}{2 \pi v} \frac{q^2}{v^2 k_0^4}
 \, \mathcal{P} \!\int_{-vk_0}^{vk_0} \! d\epsilon \, \frac{(2 \epsilon +\epsilon_q)^2}{2 \epsilon + \epsilon_q + \Omega}\,
 n_F(\epsilon+\mu) + (\Omega \to - \Omega) \, ,  \end{split}\\
\begin{split}
  & \text{Im}\, \Pi_{\eta,-\eta}(q,\Omega) = \frac{1}{4 v} \frac{(v q)^2 \Omega^2 }{(v k_0)^4}
  \frac{\sinh \frac{\Omega}{2T} }{\cosh \frac{\Omega}{2T}+ \cosh\frac{v \eta q+2\mu}{2 T }} \, ,
 \end{split}
\end{align}
where ${\cal P}$ denotes the principal value, $\epsilon = v \eta k$, and $\epsilon_q = v\eta q + 2\mu$.
Note that the real part of the backscattering polarization operator diverges  at the ultraviolet momentum
scale $k_0$ as $k_0^2$ (with the dynamical part diverging logarithmically in $k_0$) while having $k_0^{-4}$
in front of the integral, which means that the contribution of $\Pi_{\eta,-\eta}$ to ${\rm Re}\,\Pi$ is much
smaller than that of $\Pi_{\eta\eta}$, so that ${\rm Re}\,\Pi$ can be approximated (we do not directly use the
Kramers-Kronig relation for $\Pi$ anywhere in the paper) as
\begin{align}
\text{Re} \, \Pi (q,\Omega) \simeq \frac{1}{\pi v} \frac{(v q)^2}{(v q)^2 - \Omega^2}~.
\label{a10}
\end{align}
For the imaginary part of $\Pi$, we have
\begin{align}
\text{Im} \, \Pi(q,\Omega)=\frac{\Omega}{2 v} \left[\, \delta(v q-\Omega) + \delta(v q + \Omega) \,\right]
+ \frac{1}{4 v} \,\frac{(vq)^2 \Omega^2 }{(v k_0)^4} \sinh \frac{\Omega}{2T}
\left[\,\frac{1}{\cosh \frac{\Omega}{2T} + \cosh\frac{v q +2\mu }{2T}}+(q\to -q)\,\right].
\label{polarizationoperator}
\end{align}

\begin{figure}
  \centering
  \includegraphics[width=.7\textwidth]{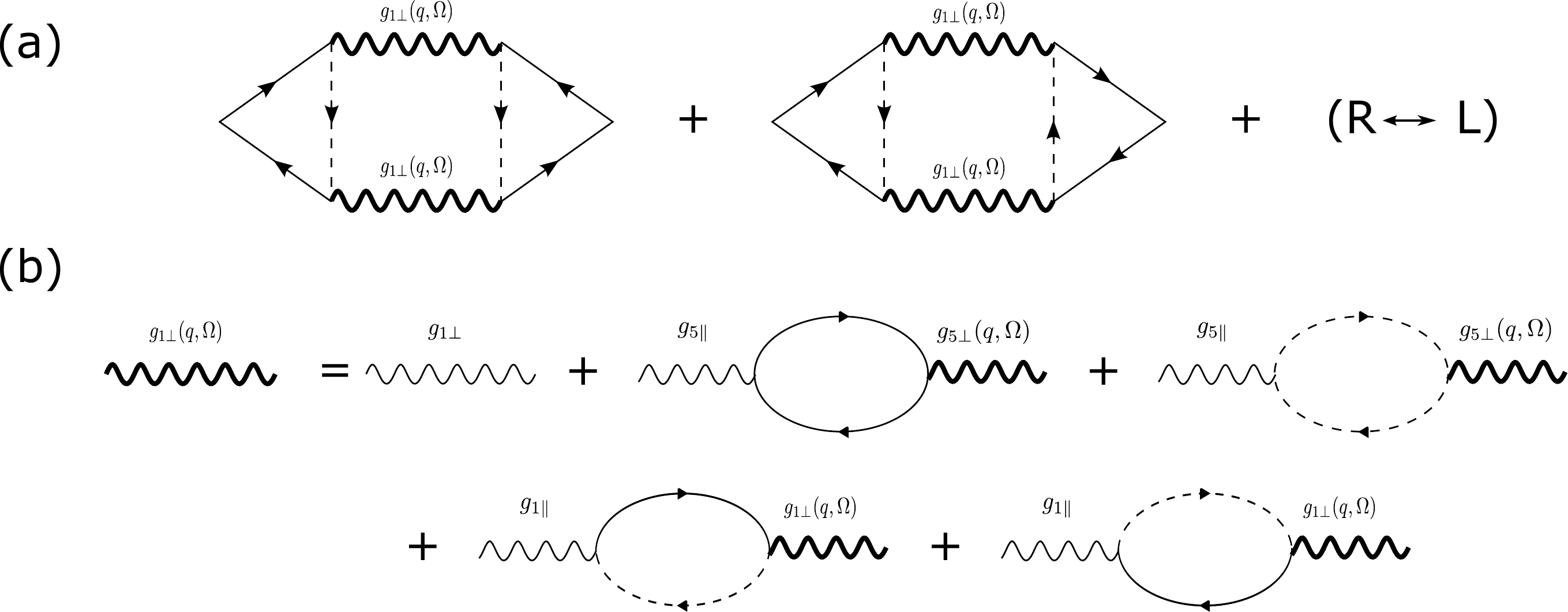}
  \caption{(a) Aslamazov-Larkin diagrams describing the lowest order contribution to drag. The solid (dotted)
  lines refer to the quasiparticle Green's functions of right (left) movers and the wiggly line denotes the dynamically
  screened RPA interaction. (b) Diagrammatic representation of the Dyson equation for $g_{1\perp}$ type interaction.
   We note that the coupling to plasmons (chiral polarization bubbles) is due to $g_5$ type interaction lines unique
   to the helical Luttinger liquid. In both (a) and (b) we have set $g_{1\parallel} = g_{3\parallel}$
   and $g_{1\perp} = g_{3\perp}$. }
  \label{Fig:dragdiagrams}
\end{figure}

We now turn to the calculation of the dynamically screened RPA interaction. The intra- and interedge components
of the interaction, $V_{11}(q,\Omega)$ and $V_{12}(q,\Omega)$, respectively, obey the Dyson equation
\begin{align}
\begin{pmatrix}V_{11} & V_{12} \\ V_{12} & V_{11} \end{pmatrix} = \begin{pmatrix}U_\parallel & U_\perp \\
U_\perp & U_\parallel \end{pmatrix} -
\begin{pmatrix}U_\parallel & U_\perp \\ U_\perp & U_\parallel \end{pmatrix} \begin{pmatrix}\Pi & 0 \\
0 & \Pi \end{pmatrix} \begin{pmatrix}V_{11} & V_{12} \\ V_{12} & V_{11} \end{pmatrix} \, ,
\label{rpa}
\end{align}
where $U_\parallel (q)$ and $U_\perp (q)$ are the bare interactions. A diagrammatic representation of the Dyson equation is presented in Fig.~\ref{Fig:dragdiagrams}. Solution to Eq.~(\ref{rpa}) reads
\begin{align}
V_{11} =& \frac{U_\parallel + \Pi \,(U_\parallel^2-U_\perp^2)}{1+2\Pi U_\parallel + \Pi^2 (U_\parallel^2-U_\perp^2)} \, ,
\\
V_{12} =& \frac{U_\perp}{1+2\Pi U_\parallel + \Pi^2 (U_\parallel^2-U_\perp^2)} \, .
\end{align}
Within the model, as discussed below Eq.~(\ref{elementsB}), we assume that $U_\parallel(q)=V_0$ is independent
of $q$ and $U_\perp (q)=V_0e^{-|q|d}$. The interedge RPA interaction $V_{12}(q,\Omega)$ can then be written as
\begin{align}
   V_{12}(q,\Omega) = \frac{1}{e^{|q| d} \left[\,V_0^{-1}+2\Pi (q,\Omega)\,\right] + 2V_0 \sinh(|q| d)\, \Pi^2 (q,\Omega)} \, .
   \label{a15}
\end{align}
Neglecting $({\rm Im}\,\Pi)^2$ compared to $({\rm Re}\,\Pi)^2$ in the real part of the term $\Pi^2$ in
the denominator of Eq.~(\ref{a15}) and using ${\rm Re}\,\Pi$ from Eq.~(\ref{a10}), $V_{12}(q,\Omega)$ reduces to
\begin{align}
   V_{12}(q,\Omega) = \frac{V_0 e^{-|q| d} [(v q)^2 - \Omega^2]^2}{(\Omega^2-\Omega_+^2) (\Omega^2-\Omega_-^2)  + 2 i ({\rm Im\,\Pi}) V_0 [(v q)^2 - \Omega^2]^2
   \left[1 + 2 e^{-|q| d} \sinh(|q|d)({\rm Re}\,\Pi) V_0 \right] } \, , \label{interedgeRPA1}
\end{align}
where the plasmon modes $\Omega_{\pm}(q)$ are obtained as the solution of the equation
\begin{align}
   [(vq)^2 - \Omega^2]^2 + 2 \alpha (vq)^2 [(vq)^2 - \Omega^2] + 2 \alpha^2 e^{-|q|d} \sinh(|q| d) (vq)^4 = 0
\end{align}
with $\alpha = V_0 /\pi v$, which gives $\Omega_\pm(q)$ in Eqs.~(\ref{plasmonpolesmaintext})--(\ref{plasmoninteractionmaintext}).

Taking the plasmon damping into account, the denominator of Eq.~(\ref{interedgeRPA1})
with the inclusion of the term proportional to ${\rm Im}\,\Pi$ is expressible as
\begin{align}
   \left\{ [\,\Omega+i \Gamma(q,\Omega)\,]^2 - \Omega_+^2(q)\right\}
   \left\{ [\,\Omega+i \Gamma (q,\Omega)\,]^2 - \Omega_-^2(q) \right\} \, .
   \label{interedgeRPAqOmega}
\end{align}
\end{widetext}
In the limit of weak damping, $|\Gamma_+ - \Gamma_- |\ll | \Omega_+(q)-\Omega_-(q)|$, where
\begin{align}
   \Gamma_\pm=\Gamma(q,\Omega_\pm)~,
\end{align}
this reduces to
\begin{align}
   \left[ (\Omega+i \Gamma_+)^2 - \Omega_+^2 \right]  \left[ (\Omega+i \Gamma_-)^2 - \Omega_-^2 \right] \, ,\label{interedgeRPA2}
\end{align}
as in Eq.~(\ref{RPAV12}).

\section{Second-order backscattering}
\label{App:Derivation of higher order backscattering processes}
In this Appendix,
we perform a real-space RG procedure using the operator product expansion~\cite{Cardy_book} (OPE)
to derive the most relevant operator generated by the backscattering term in Eq.~(\ref{bosonic model}):
\begin{align}
\begin{split}
   S_{1} =& \; \frac{g_{1\perp}}{\pi }\!  \int \! \mathrm{d} x \mathrm{d} \tau \, \cos\left(\sqrt{8\pi}\varphi_{-}\right)
   \left[\,(\partial_x\theta_+)^2-(\partial_x\theta_-)^2\,\right] \, .
\end{split}
\end{align}
The general form for an OPE for two operators $\mathcal{O}_i$ and $\mathcal{O}_j$ is
\begin{align}
\begin{split}
&:\mathcal{O}_i(\bs{r}_{\lambda,1}): :\mathcal{O}_j(\bs{r}_{\lambda,2}): \\
&= \sum_{k} \frac{c_{ijk}}{|\bs{r}_{\lambda,1}-\bs{r}_{\lambda,2}|^{\Delta_i+\Delta_j-\Delta_k}}
:\mathcal{O}_k\left( \frac{\bs{r}_{\lambda,1}+\bs{r}_{\lambda,2}}{2} \right) :
\end{split}
\end{align}
where $:\!\mathcal{O}\!:$ denotes normal ordering, $\Delta_i$ is the scaling dimension of $\mathcal{O}_i$, and
$\bs{r}_{\lambda} = (x, v_{\lambda} \tau)^T$ denotes
coordinates in space-time. The above equality does not hold on the level of operators, but it is
valid when used within the correlation functions, i.e., when the averaging is performed with another set of operators,
at a distance much larger than $|\bs{r}_1-\bs{r}_2|$ from $\bs{r}_1$ or $\bs{r}_2$.

It is convenient to introduce the complex coordinates ($\bar{z}_{\lambda}$) and ($z_{\lambda}$) as
\begin{align}
   z_{\lambda} = v_{\lambda} \tau + i x\, ,\qquad \bar{z}_{\lambda} =  v_{\lambda} \tau -i x \, ,
\end{align}
where $\tau = -i t$ is the imaginary time variable. We further
introduce the short-hand notations $1_{\lambda} \equiv (z_{\lambda,1},\bar{z}_{\lambda,1})$ and
$z_{\lambda,12} \equiv z_{\lambda,1}-z_{\lambda,2}$.
By expanding the partition function of the model defined in Eq.~(\ref{bosonic model})
in powers of $g_{1\perp}$, followed by the reexponentiation, we find the effective action~\cite{Cardy_book} to the
second order in the coupling constant,
\begin{align}
   S_{2} =  \frac{1}{2} \big[ \braket{S_{1\perp}}^2 - \braket{S_{1\perp}^2} \big] \, , \label{Seff}
\end{align}
where $\braket{\ldots}$ denotes the averaging with respect to the fixed-point action.
Within the RG procedure, we increase the short-distance cutoff $a$ at each step
by an infinitesimal amount, $a \to a' = (1+ \ell) a $, which reproduces the action, but with
renormalized coupling constants,
and may lead to the emergence of new operators.
To study the terms in the effective action, we need the time-ordered $\varphi_{\lambda} \varphi_{\lambda}$
correlation function of the $\lambda=\pm$ fields,
\begin{align}
   \braket{\varphi_{\lambda}(z_{\lambda},\bar{z}_{\lambda}) \varphi_{\lambda}(0,0)}
   = - \frac{K_{\lambda}}{4 \pi} \ln\left[\frac{|z_{\lambda}|^2 +a^2}{a^2}\right] \, .
\end{align}
The correlation function for the $\theta_{\lambda}$ fields can be obtained by using the duality relations
\begin{equation}
 K_{\lambda} \partial_{z_{\lambda}} \theta_{\lambda} = \partial_{z_{\lambda}} \varphi_{\lambda}~,
\quad K_{\lambda} \partial_{\bar{z}_{\lambda}} \theta_{\lambda} = - \partial_{\bar{z}_{\lambda}} \varphi_{\lambda}~,
\label{duality}
\end{equation}
which, similarly to the OPE, hold when used for the averages that produce the correlation functions.

The most relevant perturbation in the effective action~(\ref{Seff}) is obtained by contracting all
$\partial_x\theta$ terms for small space time distances $a<|z_{-,12}| < a'$. Using the correlation function of
the bosonic fields and the duality relations (\ref{duality}), we find the OPEs
\begin{align}
\begin{split}
   &\left[ \left(\partial_x \theta_+\right)^2 e^{i \sqrt{8 \pi} \varphi_-} \right]_1
   \left[ \left(\partial_x \theta_+\right)^2 e^{i \sqrt{8 \pi} \varphi_-} \right]_2
   \vphantom{\left(\frac{4}{L^2} \right)^{2 K_-}} \\
   \to  & \; \frac{1}{4 (\pi K_+)^2} \frac{(z_{+,12}^2+\bar{z}_{+,12}^2)^2}{(a^2 +|z_{+,12}|^2)^4}
   \left(\frac{|z_{-,12}|^2+a^2}{a^2} \right)^{2 K_-} \\
          & \; \times e^{i \sqrt{8 \pi}[ \varphi_-(1_-) + \varphi_-(2_-)]} \vphantom{\left(\frac{4}{L^2} \right)^{2 K_-}}  \, ,
\end{split}
\end{align}
and
\begin{align}
\begin{split}
   &\left[ \left(\partial_x \theta_-\right)^2 e^{i \sqrt{8 \pi} \varphi_-} \right]_1
   \left[ \left(\partial_x \theta_-\right)^2 e^{i \sqrt{8 \pi} \varphi_-} \right]_2 \\
   \to & \; \frac{1}{(4 \pi)^2}
   \Big[ \frac{2}{K_-^2} \frac{(z_{-,12}^2+\bar{z}_{-,12}^2)^2}{(a^2 +|z_{-,12}|^2)^4} + 4 \frac{(z_{-,12}+\bar{z}_{-,12})^4}{(a^2 +|z_{-,12}|^2)^4} \\
        & \;  - \frac{8}{K_-}\frac{(z_{-,12}+\bar{z}_{-,12})^2}{(a^2 +|z_{-,12}|^2)^2}\frac{z_{-,12}^2+\bar{z}_{-,12}^2}{(a^2 +|z_{-,12}|^2)^2} \Big] \\
        & \; \times  \left(\frac{ |z_{-,12}|^2+a^2}{a^2} \right)^{2 K_-} e^{i \sqrt{8 \pi}[ \varphi_-(1_-) + \varphi_-(2_-)]} \, .
   \end{split}
\end{align}
Here, we neglected less relevant terms in the OPE.
We perform the integration over the relative coordinates by introducing the polar coordinates
$z_{-,12}= r e^{-i \phi}$ and $z_{+,12}= r e^{-i \phi} + r \tilde{v} \cos \phi$
with the parameter $\tilde{v} =v_+/v_- -1= K_-/K_+-1$. The radial and angular integrations decouple
and we perform the radial integration over an infinitesimal shell $r \!\in\! (a,a')$ by setting $r= a$.
After integrating out the relative coordinates, we obtain the following contribution to the effective action:
\begin{align}
   \delta S_2 =  \frac{g_{1,\perp}^2 F(K_-,K_+) \ell }{(2 \pi)^2 v_- }
\int \! \frac{\mathrm{d} x \mathrm{d} \tau}{\pi a^2} \,  \cos[\sqrt{32 \pi} \varphi_-(x,\tau)]~,
   \label{S1action}
\end{align}
with the dimensionless function
\begin{equation}
F(K_-,K_+)= 2^{2K_-}\left[f_1(K_-,K_+) + f_2(K_-)\right]~,
\label{FKK}
\end{equation}
where
\begin{align}
\begin{split}
 f_1(x,y)  =&  4y^2
\int_0^{2\pi} \! \frac{\mathrm{d} \phi}{2\pi} \,
\frac{ \left[  (x^2+y^2) \cos^2 \phi -y^2\right]^2}
{ \left[ 2 y^2 + (x^2-y^2) \cos^2\phi \right]^4} \\
\; =& \;\frac{5x^6+45x^4y^2+7x^2y^4+7y^6}{32\sqrt{2} (x^2+y^2)^{7/2}}
\end{split}
\end{align}
and
\begin{align}
\begin{split}
 f_2(x) =& \; \frac{1}{(4 x )^2}
\int_0^{2\pi} \! \frac{\mathrm{d} \phi}{2\pi} \,\big[1-4x+6x^2 \\
\; -&
 8x(1-x)\cos 2\phi + (1-4x+2x^2)\cos 4\phi  \big] \\
\; =& \; \frac{1-4x+6 x^2}{16 x^2}~.
\end{split}
\end{align}
Importantly, the function $F(K_-, K_+)$ is nonzero for $K_\pm>0$.
We thus see that, upon renormalization, the new coupling constant is always
generated in the effective action, even if it is absent at the ultraviolet scale.
The effect of the term (\ref{S1action}) on the phase diagram of capacitively coupled
helical edge modes is discussed in Sec.~\ref{Sec:Renormalization group analysis}.

\section{Renormalization of the drag resistivity}
\label{App:Renormalization}

In this Appendix, we derive the asymptotics of the drag resistivity at $T\to 0$ for $K_->1/3$.
We assume for simplicity that the interedge interaction is weak.
To the lowest order in the interwire interaction, the dc drag resistivity can be expressed as~\cite{Pustilnik_2003,Aristov_2007}:
\begin{align}
   \rho_D = \frac{1}{e^2}\int_0^{\infty} \hspace{-.2cm} \mathrm{d} q \int_0^{\infty} \! \mathrm{d} \omega \,
   \frac{q^2 V_{12}^2(q)  }{2 K^2 k_F^2 T} \frac{\text{Im} \Pi_1(q,\omega) \text{Im} \Pi_2(q,\omega)}{\sinh^2\left( \frac{\omega}{2 T} \right)} \, ,
   \label{dragresistivity}
\end{align}
where $\text{Im} \Pi_{\sigma}(q,\omega)$ is the imaginary part of the retarded density-density
correlation function of wire $\sigma=1,2$. Here, we restrict the discussion to equal edges with the Luttinger parameter
$K_1=K_2 \equiv K$ (or, equivalently, $K_-=K_+=K$) and the plasmon velocity $v_1=v_2 \equiv v$.
The drag resistivity obtained by this conventional formula is equivalent to that obtained from the
high-frequency drag conductivity using the
kinetic equation approach \citep{Dmitriev_Gornyi_Polyakov_2012}.

We write the density operator of helical fermions by employing the expansion in Eq.~(\ref{helicalrotation}).
This yields
\begin{align}
\begin{split}
   & \rho_{\sigma}(x) = \psi_{\sigma,\uparrow}^{\dagger} \psi_{\sigma,\uparrow}^{\phantom{\dagger}}
   +  \psi_{\sigma,\downarrow}^{\dagger} \psi_{\sigma,\downarrow}^{\phantom{\dagger}} \simeq
            R_{\sigma}^{\dagger} R_{\sigma}^{} +L_{\sigma}^{\dagger} L_{\sigma}^{}  \\
            & + \frac{2 k_F}{k_0^2}
            \Big\lbrace i  \left[ (\partial_x R_{\sigma}^{\dagger} ) L_{\sigma}
            - R_{\sigma}^{\dagger} \partial_x L_{\sigma} \right] e^{-i 2 k_F x} + \text{H.c.} \Big\rbrace   \, .
            \end{split}
\end{align}
The polarization operators entering Eq.~(\ref{dragresistivity})
are calculated in the presence of the intrawire interaction which would lead to the Luttinger-liquid renormalization
of the drag resistivity, but neglecting correlations between the edges. This amounts to setting $g_{2\perp}=0$ and $g_{4\perp}=0$.
Then the quadratic part of the total Hamiltonian separates into two independent sectors in the edge basis.
In the bosonic language, the $2 k_F$-part of the density-density correlation function,
which determines the behavior of the drag resistivity at low temperatures,
can be cast in the form
\begin{align}
\begin{split}
      & \Pi^{ 2 k_F}(x,\tau) = \frac{4 k_F^2}{\pi a^2 k_0^4} e^{-i 2 k_F x} \\
      &\times  \braket{ \partial_x \theta(x,\tau) \partial_x \theta(0,0)
      e^{i \sqrt{4 \pi} [\varphi(x,\tau) - \varphi(0,0) ]}  }  + \text{H.c.}\, .
      \end{split}
\end{align}
The analytic continuation to real time and the Fourier transform to the frequency-momentum
space is standard~\cite{Giamarchi_book} and yields
\begin{align}
   \Pi^{2 k_F}(q,\omega) = \tilde{\Pi}^{2 k_F}(q+2 k_F,\omega) + \tilde{\Pi}^{2 k_F}(q-2 k_F,\omega)
\end{align}
with $\tilde{\Pi}^{2 k_F}(q,\omega)$ given by
\begin{align}
\begin{split}
      \tilde{\Pi}^{2 k_F}(q,\omega) =& - \left(\frac{k_F}{k_0} \right)^2 \frac{1}{(k_0 a)^2}
      \left( \frac{\pi a T}{v} \right)^{2K} \frac{1}{\pi^4 T^2} \\
      &\times \mathcal{K}_K\left( \frac{q v}{4 \pi T}, \frac{\omega}{4 \pi T} \right)  \, .
      \end{split}
\end{align}
Here
\begin{eqnarray}
\mathcal{K}_K\left( x, y \right) &=& \left(\frac{1}{K}+1\right) \mathcal{I}_{K+2,2}(x,y) - 2 \mathcal{I}_{K,0}(x,y)
\nonumber\\ &+&\mathcal{J}_{K+1}(x,y)
\end{eqnarray}
and we have defined the functions
\begin{align}
\begin{split}
  & \mathcal{I}_{\gamma,\delta}(x,y) = \sin(\pi \gamma) 2^{2 \gamma - \delta-2} \\
  & \times  B\left( -i (x+y) + \frac{\gamma-\delta}{2} , -\gamma +\delta +1 \right) \\
  & \times  B\left( -i  (y-x) + \frac{\gamma}{2} , -\gamma  +1 \right) + (x \to -x) \, ,
  \end{split}
\end{align}
and
\begin{align}
\begin{split}
&   \mathcal{J}_{\gamma}(x,y) = \frac{v}{(\pi T)^2} 2^{2 \gamma - 4} \sin(\pi \gamma) \\
 \times  \Big\lbrace   \Big[
                           & B\left( -i  (x+y)  + \frac{\gamma}{2} -\frac{1}{2}, -\gamma  +1 \right)  \\
                         + & B\left( -i  (x+y) + \frac{\gamma}{2} +\frac{1}{2}, -\gamma  +1 \right)
                         \Big] \\
                         \times  \Big[
                           & B\left( -i (y-x) + \frac{\gamma}{2} -\frac{1}{2}, -\gamma  +1 \right)  \\
                         + & B\left( -i (y-x) + \frac{\gamma}{2} +\frac{1}{2}, -\gamma  +1 \right)
                         \Big] \Big\rbrace \, ,\end{split}
\end{align}
where $B(x,y)$ is the Euler beta-function.
When deriving this result, we used
$$
   \int_0^{\infty} \! \mathrm{d} X \, e^{-\mu X} \sinh^{\nu}(\gamma X)
   = \frac{1}{2^{\nu+1} \gamma} B\left( \mu /2  \gamma - \nu/2, \nu+1 \right) \, ,
$$
where the identity holds as long as $\text{Re} \gamma >0$, $\text{Re} \nu > -1$ and $\text{Re} \mu > \text{Re} (\gamma \nu)$.
In our problem, there exist integrals for which the condition $\text{Re} \nu > -1$, which ensures the infrared convergence,
is not fulfilled. In that case, the integrals over time $t$ are cut off at small $t$ by $a/v$ and,
consequently, the integrals over $X$ are cut off by $\pi T a/v$.

For $T \to 0$, the function $\text{Im}\; \mathcal{K}\left( [q-2 k_F]/ 4 \pi T, \omega/ 4 \pi T \right)$ is strongly peaked
around $q = 2 k_F$ with a width of the peak of the order of $T/v$. Therefore, we can neglect the term
$\text{Im}\; \mathcal{K}_K\left( [q+2 k_F]/ 4 \pi T, \omega/ 4 \pi T \right)$
in the integral over positive momenta in Eq.~(\ref{dragresistivity}). Then, we find
\begin{align}
   \rho_D  \sim  I_K \frac{\left[V_{12}(2k_F)\right]^2}{v^2} \left(\frac{k_F}{k_0} \right)^4 \frac{T}{(k_0 a)^4}
   \left( \frac{\pi a T}{u} \right)^{4K} \, ,
   \label{Luttingerdrag1}
\end{align}
where
$$I_K= \int_0^{\infty} \! \mathrm{d} \Omega \, \frac{[\,\text{Im}\, \mathcal{K}_K(0,\Omega/4 \pi)\,]^2}{\sinh^2(\Omega/2)}$$
with $\Omega = \omega/T$.
As discussed in the main text, the natural ultraviolet cutoff here is provided by the distance between the edges, $a\sim d$.
The parametric dependence of the drag resistivity obtained by means of bosonization reproduces in the limit $K \to 1$ the result (\ref{dcdrag1})
of the kinetic-equation analysis.

\bibliography{database}

\end{document}